\documentclass{paper}

\usepackage[a4paper]{geometry}
\usepackage{graphicx}
\usepackage[textwidth=8em,textsize=small]{todonotes}
\usepackage{amsmath}
\usepackage{natbib}
\usepackage{subfigure}
\usepackage{amsfonts}
\usepackage{amssymb}
\usepackage{mathtools}
\usepackage{multirow}
\usepackage{appendix}
\usepackage{indentfirst}
\usepackage{hyperref}
\usepackage[utf8]{inputenc}
\usepackage{bm}
\usepackage{amsmath}
\usepackage{float}
\newcommand{\myD}{{\mathcal{D}}}
\newcommand{\myL}{{\mathcal{L}}}
\newcommand{\myI}{{\mathcal{I}}}
\newcommand{\bftheta}{{\boldsymbol{\theta}}}

\usepackage{array}
\usepackage{wrapfig}
\usepackage{tabularx}

\title{A decision support system for addressing food security in the UK}
\author{Martine J. Barons
, Tha{\'i}s C. O. Fonseca, Andy Davis and Jim Q. Smith} 
\begin{document}
\date{\today}
\maketitle

\begin{abstract}
This paper presents an integrating decision support system to model food security in the UK.  In ever-larger dynamic systems, such as the food system, it is increasingly difficult for decision makers to effectively account for all the variables within the system that may influence the outcomes of interest under enactments of various candidate policies.  Each of the influencing variables are likely, themselves, to be dynamic sub-systems with expert domains supported by sophisticated probabilistic models. Recent  increases  in  food  poverty  the  UK raised the questions about the main drivers to food insecurity, how this may be changing over time and how evidence can be used in evaluating policy for decision support. In this context, an integrating decision support system is proposed for household food security to allow  decision makers to compare several candidate policies which may affect the outcome of food insecurity at household level.

\end{abstract}

\section{Introduction}

This paper gives a proof of concept practical application of the recently developed statistical integrating decision support system (IDSS) paradigm.  An IDSS is developed for policymakers concerned with deciding between candidate policies designed to ameliorate household food insecurity within the UK context of rising food charity use.

\subsection {Food Security}

Food security exists when all people, at all times, have physical and economic access to sufficient, safe and nutritious food to meet their dietary needs and food preferences for an active and healthy life \citep{FAO1996}. Missing meals and changing diet is a common response to food insecurity, and the latter may persist over extended periods, leading to adverse health effects, especially in children \citep{Seligman2010}. 
Food insecurity can result in an increased risk of death or illness from stunting, wasting, weakened responses to infection, diabetes, cardiovascular diseases, some cancers, food-borne disease and mental ill health, via insufficient quantity, poor nutritional quality of food, contaminated foods, or social exclusion \cite{Friel2015}.  
Rising food insecurity has been strongly associated not just with malnutrition, but with sustained deterioration of mental health, inability to manage chronic disease, and worse child health \citep{Loopstra2015a, LoopstraThesis2014}. 
Food insecurity is associated with hypertension and hyperlipidemia which are cardiovascular risk factors.  It is also associated with poor glycaemic control in those with diabetes, whose additional medical expenses exacerbate their food insecurity \citep{NHANESLee2019}. Food insecurity has been found to affect school children’s academic performance, weight gain, and social skills \citep{Faught2017}.  Whilst obesity is more prevalent among food-insecure women,  controlling for BMI did not attenuate the association of food insecurity and chronic disease \citep{Pan2012}. 

\subsection{The UK picture}

The recent increases in food insecurity the UK is well known through the much-publicised increase in the uptake of humanitarian aid, principally through food banks and their corresponding increase in number \citep{Loopstra2015}.  As a nation, the UK is wealthy and one of the world’s most food secure; in 2017 it was 3rd of 113, just after Ireland and the USA \citep{GlobalFoodSecurityIndex} but by December 2019 has declined to 17th place. 
In 2013, a letter published in the BMJ \citep{Taylor-Robinson2013} on the rise of food poverty in the UK alerted readers to the fact that the number of malnutrition-related admissions to hospital had doubled since 2008/9. 
When food parcel distribution by the Trussell Trust exceeded one million in 2014/15, this was interpreted by some as evidence that the UK government is not fulfilling its legal duty under the International Covenant on Economic, Social and Cultural Rights \citep{ICESCR} to take appropriate steps to realise the right of everyone to be free of hunger. Year ending March 2019 more than 1.6 million parcels were distributed, and in the six months to September 2019, the number of parcels had risen by 23\% on the previous year \citep{Trussell2019}.
Persistent  and widespread low pay, the proliferation of zero-hours contracts and rising living costs, especially food prices, have been suggested as contributory factors for the increase in food insecurity, and the health consequences of inadequate diets have also been raised by health professionals \citep{CSI13}. 
Relative to other advanced western economies, Britain had higher general inflation, higher food, fuel and housing price inflation, lower growth in wages in the years immediately following the 2008 global financial crisis.  The UK also has a history of very large numbers of very low paid employees; many of those accessing food banks are in work \citep{APPG2014}. 
For many years, the exact scale of the problem in the UK was unknown. This was because there was no systematic, national assessment of the numbers of households experiencing food insecurity, but only small-scale studies \citep{Pilgrim2012}, \citep{Tingay2003}. However, from 2016, the Food Standards Agency included the Adult Food Security Module of the USDA Household Food Security Survey (HFSS) \citep{HFSS_adult} 
in the bi-annual Food and You Survey. The HFSS contains 10 items for households without children and 18 items for households with children (age 0 - 17) to assess their experiences over the last 12 months. The HFSS classifies households as being food insecure when the respondent reports three or more food insecure conditions and as very low food security category if at least one member experienced reduced food intake or if insufficient resources for food disrupted eating patterns. The latest UK survey, Wave 5 (2018) \citep{FoodandYou5}, found that  80\% of respondents lived in households with high food security, 10\% in households classified as marginally food secure, and 10\% reported living in household with low or very low food security. There is more food insecurity amongst families with children: those who lived with children under the age of 16 were less likely than those with no children to have high levels of food security (70\% compared with 84\%). Employment and income are key determinants of food security. Nearly a quarter (23\%) of unemployed people lived in households with very low food security, compared to 4\% of those in work. In the lowest income group, 59\% of households had high food security, increasing with income to 93\% in the highest income households. In households in the lowest income groups, 13\% had very low food security (compared with less than 1\% of those in the highest income households). 

\subsection {Comparison with USA and Canada}
Like the UK, USA and Canada, are  wealthy nations with significant household food insecurity.  In contrast to the UK, the USA and Canada have undertaken regular monitoring of household food security over many years through the HFSS module within regular household surveys \citep{Canada2016}. This means that research on determinants and rates of food insecurity over time is more advanced and detailed in USA and Canada than in the UK.\\
The USA and Canada are similar to the UK in their profiles of poverty and types of government, which allows us to draw on their research where UK data and evidence is sparse.  

In 2017-18, and the UK absolute poverty rate was 19.0\%, ranging from 26.5\% among children to 13.5\% among pensioners \citep{UKPoverty2019}.
In the USA, the official poverty rate in 2018 was 11.8\%, for children under age 18 it was 16.2\%, for people aged 18 to 64, 10.7\% and  for people aged 65+, 9.7\% \citep{USAPoverty2019}.  
In Canada, the official poverty rate is 9.5\% overall and 9.0\% for children.
3.9\% of seniors were living in poverty in 2017 \citep{CanadaPoverty2017}, although the Market Basket Measure has been criticised for omitting housing and childcare costs.  The Canadian Low Income measure, 50\% of median income, adjusted for family size, was 12.9\% in 2017 on an after-tax basis.
 In 2018 in the USA, 11.1\% of households  were food insecure and 4.3\% had very low food security. In Canada it was 12.3\% in 2011, the latest figures available, with 2.5\% of households with very low food security. \citep{LoopstraThesis2014,Tarasuk2010}

\begin{table}
\caption{Poverty measures across three countries. UK  absolute poverty rate  measures the fraction of population with household income below 60\% of median income in 2010–11, updated by the Consumer Prices Index. USA  Census Bureau uses a set of dollar value thresholds that vary by family size and composition to determine poverty. Canada uses the Market Basket Measure, the concept of an individual or family not having enough income to afford the cost of a basket of goods and services.\label{table:Pov}}
\centering
\begin{tabular}{|l c c c|} 
 \hline                              & UK     & USA   & Canada  \\  
 \hline
 Overall                        & 19.0\%      & 11.8\%    & 9.5\% \\ 
 Child Poverty                  & 26.5\%    & 16.2\%    & 9.0\% \\
 Working adults with no children & 16.4\%   & --        & -- \\
 Adults 18-64                   & --        & 10.7\%    & -- \\
 Pensioners                     & 13.5\%    & 9.7\%     & 3.9\% \\
 Food security low (very low)   & 10.0\%      & 11.1\% (4.3\%) & 12.3\% (2.5\%)\\  
 \hline
\end{tabular}
\end{table}

\subsection{Need for decision support}

There is a need to gather what information does exist for the UK in order to ascertain the principal drivers of household food security to support policy-makers to design policy to tackle food security and to evaluate other policies which may impact on food security.\\
In ever-larger dynamic systems, such as the food security, it is increasingly difficult for decision makers to effectively account for all the variables within the system that may influence the outcomes of interest under enactments of various given policies. In particular, government policies on welfare, farming, the environment, employment, health, etc. all have an impact on food security at various levels. Each of the influencing variables are likely, themselves, to be dynamic sub-systems with domain expertise, often supported by sophisticated probabilistic models. Within the food system, examples of these are medium to long range weather forecasting which influences food supply which might be large numerical models, and economic models such as autoregressive or moving average  which estimate the behaviour of global markets and prices under various plausible scenarios. The emerging crisis in the UK is not merely a matter for charity, but of great concern to policymakers, who are legally and morally obligated to act, but may lack recent experience in dealing with needs of this kind and scale, and so require decision support. This paper proposes an integrating decision support system (IDSS) \citep{SmithBarons2015, Barons2017} for household food security in the UK. The IDSS is a computer-based tool which integrates uncertainties of different parts of a complex system and addresses the decision problem as a whole.

\subsection{Practical considerations}
In \cite{Barons2017}, we detail the iterative manner of the development of an IDSS with its decision-makers and expert panels. Before the elicitation starts it is always necessary to do some preparatory work. With the help of various domain experts, the analyst will need to trawl any relevant literature and check which hypotheses found there might still be current.
We repeatedly review the qualitative structure of the IDSS in light of the more profound understanding of the process acquired through more recent elicitation. This modification and improvement continues until the decision centre is content that the structure is requisite \citep{Phillips1984}. Since the process of model elicitation is an iterative one, it is often wise to begin with some simple utility measures, proceed with an initial structural model elicitation, and then to revisit the initial list of attributes of the utility;  detailed exploration of the science, economics or sociology can prompt the decision centre to become fully aware of the suitability of certain types of utility attribute measures. By focusing
the centre and its expert panels on those issues that really impact on final
outcomes we can vastly reduce the scope of a potentially enormous model; only those features that might be critical in helping to discriminate between the potential effectiveness of one candidate policy against another are required. If there is strong disagreement about whether or not a dependency exists in the system then we assume initially that a dependency does exist, except where the consensus is the its effect is weak. Further iterations of the model building process usually clarify the understanding, and if not, a sensitivity analysis can usually distinguish a meaningful inclusion form others. The decision centre also need to decide what time step is the most natural one to use for the purposes of  the specific IDSS. This choice depends on the speed of the process, how relevant data is routinely collected on some of the components, and some technical acyclicity assumptions that are typically known only
to the decision analysts. There may be conflict between the granularity of  informing economic models of the process, sample survey regularity, and the needs of the system. The granularity needed is driven by the granularity of the attributes of the utility. In addition, decision analysts need to match precisely the outputs of a donating panel with the requirements of a receiving panel. When these do not naturally align, then some translation, possibly a bespoke model,  may be needed between them. When expert panels design their own systems, sometimes the internal structure of one component can share variables with the internal structure of another. So, for example, flooding could disrupt both the production of food and its distribution and yet these might be forecast using different components. In such cases, the coherence of the system will be lost and the most efficient way to ensure ongoing coherence is to separate out the shared variables and ask the panels concerned to take as inputs, probability distributions from the expert panel in the shared variable, flood risk. One element of these IDSS systems is the way they can appropriately handle uncertainties associated with various modules. This is vital to reliable decision making. For example if the inputs from one module are very speculative - and so have a high variance - Then policies that work well over a wide range of such inputs will - under the sorts of risk averse decisions we have here - ted to be preferred to ones whose efficacy is very sensitive to such inputs.  That is why we need conditional inputs to communicate such uncertainties. 

\section{Integrating decision support systems\label{DDSS}}

 Integrating Decision Support systems are introduced in \cite{SmithBarons2015} and \cite{Smith2016} and briefly reviewed in section 2.1.  The IDSS aids decision makers in the understanding of a problem by providing a clear evaluation and comparison of the possible options available. It combines expert judgement with data for each subsystem resulting in a full inferential procedure able to represent complex systems. However, decision support systems often require sophisticated architectures and algorithms to calculate the outputs needed by the decision-makers to inform policy selection when the system is composed of many multi-faceted stochastic processes. There is currently no  generic framework or software  which is capable of faithfully expressing underlying processes for the scale of problems under consideration here, nor sufficiently focused to make calculations quickly enough for practical use in a dynamic, changing environment.  

In this application, the framework knitting together the different component subsystems in the IDSS is the dynamical Bayesian Network \citep{West97}. 
In particular, the model can be seen as a multi-regression dynamic model (MDM) \citep{Queen1993}. Here this framework is extended to allow variances to vary stochastically over time. The assumed approach is suitable because regression models are well understood but we need to allow for the fact that within this application regression coefficients can drift in time.  The dynamical model also allows for separability of the different components of the series. A simulation algorithm is developed which enables decision making to be fast and dynamical over time even for a large system with many dependent variables and time points with nonlinear characteristics. Using the MDM, we can model shocks to the system within the given framework by introducing change point. This sort of property is exploited in the brain imaging \citep{Costa2019}.
 Within each of the expert panels lies a complex sub-network of variables. We seem to a BN/DBN for all the modules since these are a very well developed method used in main analogous applications and have supporting software easily available.   
In Section \ref{DDSS2}, the integrating decision support system methodology is briefly reviewed.
Section \ref{IDSSFood} details the model and variables used for utility computation in the context of food security in the UK. Then Section \ref{Out} presents the outputs and policy evaluation for the food security system. We end the paper with a short discussion of our findings and the planned next steps in this research programme.

\subsection{Technical underpinning \label{DDSS2}}

In this section, we briefly review these recent methodological developments to support inference for decision support as they apply here. Full details and proofs are provided in  \citep{Smith2016}.

Consider a vector of random variables relevant to the system 
$\mathbf{Y}=(Y_1,\ldots, Y_n)$. 
Typically, there are expert panels with expertise in particular aspects of the multivariate problem. The most appropriate expert panels for each sub-system are identified, each sub-panel will defer to the others, adopting their models, reasoning and evaluations as the most appropriate domain experts. Each expert panel, $G_i$, is responsible for a subvector $\mathbf{Y_{B_i}}$ of $\mathbf{Y}$, with $B_1,\ldots,B_m$ a partition of $1,\ldots,n$. The multivariate problem is then decomposed in sub-models. The joint model thus accommodates the diversity of information coming from the different component models and deals robustly with the intrinsic uncertainty in these sub-models. 

Decisions $d\in \mathcal{D}$ will be taken by a decision maker (DM) where $\mathcal{D}$ represents the set of all policy options that it plans to consider. In the context of large problems like this, the decision-maker is often a centre composed of several individuals.  These individuals are henceforth assumed to want to work together constructively and collaboratively supported by using a probabilistic decision tool that can provide a benchmark evaluation of $d\in \mathcal{D}$ the underlying processes that drive the dynamics of the unfolding scenario. However, to use the Bayesian paradigm, we would like to assume that this centre will strive to act an a single rational person would when that person is the owner of the beliefs expressed in the system and so the need for coherence is satisfied. The DM receives information from each panel and reaches a conclusion that depends on a reward function $R(\mathbf{Y},d)$, $\mathbf{Y}\in R_Y$, $d\in \myD$. For this level of coherence, we must be able to configure the panels and their relationships so that certain assumptions are satisfied.  Below we briefly outline what these assumptions need to be.  More generic descriptions can be found in \citep{Smith2016}.

We introduce some notation: For each $i=1, \ldots,m$ let the subvector $\mathbf{Y_{B_i}}$ be delivered by $G_i$ depend on a function $\myL_i(\mathbf{Y}_{B_i})$.  
Each panel $G_i$ provides a model $\mathbf{Y}_{B_i}\mid \myL_i(\mathbf{Y}_{B_i}),\boldsymbol{\theta}_{B_i},d$, and prior information about $\boldsymbol{\theta}_{B_i}$. Each panel $G_i$ will deliver summaries denoted by $S_i^y(\myL(Y),d)$ which are expectations of functions of $Y$ conditional on the values of $\myL(Y)$ for each decision $d\in \myD$. Let $U(R(\mathbf{Y},d))$ be the utility function for decision $d\in \mathcal{D}$. Our main goal is to compute the expected utilities $\{\bar U(d):\; d\in \myD\}$ which represents the expected utilities of a decision maker.

To be formally valid, any IDSS must respect a set of common knowledge assumptions shared by all panels and which comprises the union of the utility, policy and structural consensus, described as follows.

\begin{enumerate}
\item {\bf Structural consensus:} The structural consensus requires that all the experts agree, in a transparent and understandable manner, the qualitative structure of the problem in terms of how different features relate to one another and how the future might unfold within the system. Formally, these can be couched in terms of sets of irrelevance statements. We propose such a structure in \ref{UKDBN}. There needs to be an agreed narrative of what might happen within each component of the system, based on best evidence.  Also for each component, there needs to be a quantitative evaluation of how the critical variables might be affected by the developing environment when appropriate mitigating policies are applied.  Where there are agreed sets of irrelevance statements, and the semigraphoid axioms are assumed to hold \citep{Smith2010}, these can be used to populate the common knowledge framework belonging to a decision centre.

\item {\bf Utility consensus:} requires all to agree \emph{a priori} on the class of utility functions supported by the IDSS and the types of preferential independence across its various attributes it will need to entertain (such as value independence, mutually utility independent attributes \citep{Keeney1993} and more sophisticated versions, see \cite{Leonelli2015}. Sections \ref{UF1} and \ref{UF2} give details of the multiattribute utility, its measurement and rationale. 

\item {\bf Policy consensus:}  must be sufficiently rich to contain a set of policies that might be adopted and an appropriate utility structure on which the efficacy of these different policies might be scrutinised. 

\item {\bf Adequate:} An adequate IDSS  will be able to unambiguously calculate expected utility score for each policy that might be adopted on the basis of the panels' inputs; if it has this property the IDSS is called adequate. Note that it should be immediate from the formulae of a given probabilistic composition to calculate these expectations whether or not the system is adequate (see \cite{Smith2016} for an illustrative example).

\item {\bf Sound:} A sound IDSS is one which is both adequate and allows the decision-maker, by adopting the structural consensus, to admit coherently all the underlying beliefs about a domain overseen by a panel as her own, and so accept the summary statistics donated by the panels to the IDSS.

\item {\bf Distributive:} For such a system to be formal and functional, each component panel can reason autonomously about those parts of the system they oversee and the centre can legitimately adopt their delivered judgements as its own.  The semigraphoid axioms provide  means to satisfy this requirement and panel autonomy liberates each panel of domain experts to produce their quantitative domain knowledge in the way most appropriate for their domain and using their own choice of probability models.  They can update their beliefs through any models they might be using and continually refine their inputs to the system without disrupting the agreed overarching structure and its quantitative narrative.

\item {\bf Separately informed:} An essential condition for panel autonomy is that panel are separately informed.  This requirement can be subdivided within a Bayesian framework into two conditions - prior panel independence and separable likelihood - using the usual properties of conditional independence.  The first of these is a straightforward generalisation of the global independence assumption within Bayesian inference \citep{Cowell1999}.  The second, the assumption that the collection of data sets gives a likelihood that separates over subvectors of panel parameters, is far from automatic and is almost always violated when there are unobserved confounders or missing data. In such circumstances, one approach is to devise appropriate approximations.

\item {\bf Admissibility protocols:} Another approach is to  impose an admissibility protocol on the information used to make inferences within the system, analogous to quality of evidence rules within Cochrane Database of Systematic Reviews.  When data is derived from well-designed experiments, randomisation and conditioning often leads to a likelihood which is a function only of its own parameters, so trivially separates. When there is a consensus that a quantitative causal structure is a \emph{causal} Bayesian network, dynamic Bayesian network, chain event graph or multiprocess model and the IDSS is sound (delegable, separately informed and adequate), then the IDSS remains sound  under a likelihood composed of ancestral sampling experiments and observational sampling \citep{Smith1997}.


\item {\bf Transparent:} In such a distributive framework, any query made by another panellist or an external auditor can be referred to the expert panel donating the summaries in question which can provide a detailed explanation of its statistical models, data, expert judgements and other factors informing how its evaluation have been arrived at and why the judgements expressed are appropriate.


\end{enumerate}

For a distributive IDSS, the question then becomes precisely which information each of the panels needs to donate about their areas of expertise for the maximum utility scores to be calculated. Provided that the utility function is in an appropriate polynomial form, each panel need deliver only a short vector of conditional moments and not entire distributions because this type of overarching framework embeds collections of conditional independences allowing the use of tower rule recurrences \citep{Leonelli2015}.  This facilitates fast calculations and propagation algorithms to be embedded within the customised IDSS for timely decision-making.  In such a system, individual panels can easily and quickly perform prior to posterior analyses to update the information they donate when relevant new information comes to light and this can be propagated to update the expected utility scores; this quality is especially useful within decision support for an emergency, but in any circumstances represents a huge efficiency gain over having to rebuild and re-parameterise a large model. There are a number of frameworks which satisfy the requirements of the IDSS properties, including staged trees, Bayesian Networks, Chain graphs, Multiregression dynamic models and uncoupled dynamic BNs.

The paradigm outlined here will be illustrated throughout the remainder of the paper through a proof of concept application to an IDSS for government policy for household food security in the UK, using a Bayesian network as the overarching framework. 

\subsection{BN and Dynamical BN\label{subBN}}

Bayesian networks (BNs) and their dynamic analogues are particularly suited to the role of decision support as they represent the state of the world as a set of variables and model the probabilistic dependencies between the variables.  They are able to build in the knowledge of domain experts, provide a narrative for the system and can be transparently and coherently revised as the domain changes. 

A Bayesian network is formally defined as a directed acyclic graph (DAG) together with a set of conditional independence statements having  the form A is independent of B given C written $A \perp B|C$. They are a simple and convenient way of representing a factorisation of a joint probability density function of a vector of random variables $\mathbf{Y}=(Y_1, Y_2, \ldots, Y_n)$. Each node has a conditional probability distribution, which in the case of discrete variables will be conditional probability tables (CPTs). 
In this model, $\myL_i(\mathbf{Y}_{B_i})=\mathbf{Y}_{\Pi_i}$, with $\Pi_i$ the indices of parents of $Y_i$. The joint density of $\mathbf{Y}$ may be written as
$$f(\mathbf{y}\mid d)=\prod_{i\in [n]}f_i(y_{B_i}\mid y_{\Pi_{B_i}},d).$$

Assume that $U(R(\mathbf{Y},d))=\sum_{i\in [m]} k_i \; U_i(R_i(\mathbf{Y}_{B_i},d))$. Thus the expected utility is given by 
$\bar U(d)=\sum_{i\in [n]} k_i \; \bar U_i(d\mid y_{\Pi_i})$, with
$$\bar U_i(d\mid y_{\Pi_i})=\int_{\Theta_{B_i}}\int_{R_{y_{B_i}}} U_i(R_i(y_{B_i},d))\; f_i(y_{B_i}\mid y_{\Pi_i},\theta_{B_i},d)\; \pi_i(\theta_{B_i}\mid d) dy_{B_i}d\theta_{B_i}.$$

Dynamic Bayesian networks are able to accommodate systems which change over time \citep{Dean1990}. DBNs are a series of BNs created for different units of time, each BN called a time slice. The time slices are connected through temporal links to form the full model. 
DBNs can be unfolded in time to accommodate the probabilistic dependencies of the variables within and between time steps. 
It is usually assumed that the configuration of the BN does not change over time, i.e. the dependencies between variables are static.

Consider the general setting such that
\begin{equation}
    \mathbf{Y}_{it} \perp \mathbf{Y}_{Q_i}^ t\mid \mathbf{Y}_{\Pi_i}^ t,\mathbf{Y}_i^{t-1},\; i=1,\ldots, n,
\end{equation}
with $\{\mathbf{Y}_t:\; t=1,\ldots,T\}$ a multivariate time series composing a DAG whose vertices are univariate processes and $\Pi_i $ the index parent set of $Y_{it}$ and $\mathbf{Y}_i^ t=(Y_{i1},\ldots,Y_{it})'$ the historical data. Thus,  the model assumes that each variable at time t depends on its own past series, the past series of its parents and the value of its parents at time t. This results in the joint density function
\begin{equation}
    f(\mathbf{y})=\prod_{t=1}^{T}\prod_{i=1}^{n}f_{i,t}(y_{it}\mid y_{\Pi_i}^ t,y_i^{t-1}).
\end{equation}
The observation and system equations are defined as
\begin{eqnarray}\nonumber
Y_{it} & = & F_{it} \theta_{it} + \epsilon_{it},\\ \nonumber
\theta_{it} & = & G_{it} \theta_{i,t-1} + \omega_{it},
\end{eqnarray}
with $\epsilon_{it}\sim N[0,V_{it}]$ and $\omega_{it}\sim N[0,W_{it}]$. The errors are assumed to be independent of each other and through time and $F_{it}$, $G_{it}$ are assumed to be known. Given the initial information, $\theta_{i0}\mid \myI_0\sim N[m_{i0},C_{i0}]$. 
The parameters $\theta_{it}$, $i=1,\ldots,n$ may be updated independently given the observations at time $t$. Conditional forecasts may also be obtained independently. These results are proved in \cite{Queen1993} assuming Gaussian distributions for the error terms. The predictive density is given by
\begin{eqnarray}\nonumber
    f(\mathbf{y}_t\mid \mathbf{y}^ {t-1})& = & \int_{\Theta} f(\mathbf{y}_t\mid \mathbf{y}^ {t-1},\theta_t)\;  \pi(\theta_t\mid \mathbf{y}^ {t-1})\; d\theta_t\\ \nonumber
    & = & \prod_{i=1}^{n}\int_{\Theta_i} g_{it}(y_{it}\mid y_{\Pi_i}^t,y_i^ {t-1},\theta_{it})\pi_i(\theta_{it}\mid y_{\Pi_i}^{t-1},y_i^ {t-1})d\theta_{it}.
\end{eqnarray}

Let $\textbf{D}_{t}=(\mathbf{y}_t,\textbf{D}_{t-1})$ be the information available at time t. Inference about $\bftheta_t$ is based on Forward filtering equations to obtain posterior moments at time $t$. 

\begin{itemize}
\item[--] Posterior distribution at time $t-1$: $
\bftheta_{i,t-1}\mid \textbf{D}_{t-1} \sim N[\textbf{m}_{i,t-1},C_{i,t-1}]; 
$

\item[--] Prior distribution at time $t$: $
\bftheta_{it}|\textbf{D}_{t-1} \sim N[\textbf{a}_{it},R_{it}],$
\noindent
with $\textbf{a}_{it} = G_{it}\textbf{m}_{i,t-1}$ and $R_{it}=G_{it}C_{i,t-1}G'_{it}+W_{it} ;$

\item[--] One step ahead prediction:
$ \textbf{y}_{it}\mid  \textbf{y}_{\Pi_i,t}, \textbf{D}_{t-1} \sim N[\textbf{f}_{it},Q_{it}], $
\noindent
with $\textbf{f}_{it} = \textbf{F}'_{it}\textbf{a}_{it}$ and $Q_{it}=\textbf{F}'_{it}R_{it}\textbf{F}_{it}+V_{it}$;

\item[--] Posterior distribution at time $t$: $\bftheta_{it}\mid \textbf{D}_{t} \sim N[\textbf{m}_{it},C_{it}],$
\noindent
with $\textbf{m}_{it} = \textbf{a}_{it}+\textbf{A}_{it}\textbf{e}_{it}$ and $C_{it}=R_{it}-A_{it}Q_{it}A'_{it}$ and $\textbf{e}_{it}=\textbf{y}_{it}-\textbf{f}_{it}$, $A_{it}=R_{it}\textbf{F}_{it}Q_{it}^{-1}$.
\end{itemize}

If data is observed from time $1$ to $T$ then backward smoothing may be used to obtain the posterior moments of $\theta_{it}\mid D_T$, $t=1,\ldots,T$. Thus,
$$\theta_{it}|\theta_{i,t+1},\textbf{D}_T \sim N(\textbf{h}_{it},H_{it}),$$ 
with $\textbf{h}_{it} = \textbf{m}_{it}+C_{it} G'_{i,t+1}R_{i,t+1}^{-1}(\theta_{i,t+1}-\textbf{a}_{i,t+1})$, $H_{it} = C_{it} - C_{it}G'_{i,t+1}R_{i,t+1}^{-1}G_{i,t+1}C_{it}$ and $\textbf{h}_{iT} = \textbf{m}_{iT}$ e $H_{iT} = C_{iT}$, the initial values. \\


The variance evolution follows \cite{West97} which define $V_{it}=V/\phi_{it}$ and $\phi_{i,t-1}\mid D_{t-1}\sim G(n_{i,t-1}/2,d_{i,t-1}/2)$. The gamma evolution model is given by
$$\phi_{it}\mid D_{t-1}\sim Gamma(\delta_i n_{i,t-1}/2,\delta_i d_{i,t-1}/2),$$
with $\delta_i \in (0,1)$ being the discount factors. 
The posterior distribution at time $t$ is obtained analytically as
$\phi_{it}\mid D_t \sim Gamma(n_{it}/2,d_{it}/2)$ with $n_{it}=\delta_i n_{i,t-1}+1$ and $d_{it}=\delta_i d_{i,t-1}+S_{i,t-1}e_{it}'Q_{it}^{-1}e_{it}$, with $S_{i,t-1}= d_{i,t-1}/ n_{i,t-1}$. This conjugacy results in closed-form recurrence updating equations for this variance model. 

\subsection{Expected utility computation and scenario evaluation}

Suppose that $\theta_{1:T}$ was simulated using the Forward filtering and backwards sampling algorithm as described in subsection \ref{subBN}. The predictive posterior distribution for a replicated observation $\tilde y$ is given by

\begin{eqnarray}\nonumber
    f(\mathbf{\tilde y}_{t}\mid \mathbf{y}^ {t})& = & \int_{\Theta} f(\mathbf{\tilde y}_t\mid \mathbf{y}^ {t},\theta_t)\;  \pi(\theta_t\mid \mathbf{y}^ {t})\; d\theta_t\\ \nonumber
    & = & \prod_{i=1}^{n}\int_{\Theta_i} g_{it}(\tilde y_{it}\mid \tilde y_{\Pi_i}^t,\tilde y_i^ {t-1},\theta_{it})\pi_i(\theta_{it}\mid y_{\Pi_i}^{t},y_i^ {t})d\theta_{it}.
\end{eqnarray}

The predictive distribution of a new observation $\tilde y_{it}$ may be obtained by simulating from $g_{it}(\cdot\mid \tilde y_{\Pi_i}^t,\tilde y_i^ {t-1},\theta_{it})$. If $U(\mathbf{\tilde{y}}_{t},d)$ are linear functions of $\mathbf{\tilde y}_t$ the expected utilities may be computed analytically using chain rules of conditional probabilities. If $U(\mathbf{\tilde{y}}_{t},d)$ is a nonlinear function of $\mathbf{\tilde{y}}_{t}$ then expected values are computed by Monte Carlo integration \citep{Rob04}. Note that some ordering in computing expectations need to be followed, starting from the variables such that $\mathcal{L}_i(\mathbf{Y}_{it})=\emptyset$, their descendants and so on.

In addition, the types of overarching descriptions suitable for these applications must be rich enough to explore both the effects of shocks to the system and the application of policies.  These can be conveniently modelled through chains of causal relationships, where causal means that there is an implicit partial order to the objects in the system and we assume that the joint distributions of variables not downstream of a controlled variable remain unaffected by that control.  The downstream variables are affected in response to a  controlled variable in the same way as if the controlled variable had simply taken that value. This is the assumption underlying designed experiments.

\section{IDSS: UK Food security\label{IDSSFood}}

\subsection{Utility function elicitation}
\label{UF1}
In every decision support scenario, it is essential to clarify the goals of the decision-maker (DM).  Support for household food security is provided in the UK context through Local government, typically city or county councils through their financial inclusion and child poverty policies. The goal of a city or county council in the UK is to fulfil their statutory obligations to the satisfaction of central government.  Whenever possible, they wish to go beyond mere compliance and continually improve the lives of the citizens within their geographic region, with a special focus on improving the circumstances of the most disadvantaged.  

In order to construct an IDSS for food security, the next step is to define the utility function and develop a suitable mathematical form for it. One requirement of the attributes of a utility function is that they must be measurable; it must be possible to say whether an event has happened or a threshold has been reached. One candidate measure of household food security would be data from food bank charities. However, studies have shown that food bank use is not a good measure of food poverty \citep{Tarasuk2009,USDA2016}. In the absence of a direct measure of household food security in the UK, the decision-maker needs a good proxy in order to construct a suitable Utility function. Council officers identified the variables: education, health and social unrest as suitable attributes of a utility. 

In constructing a utility function based on these attributes, it appeared appropriate to assume value independence \citep{Keeney1993}. Let $Z_1=$measures of education, $Z_2=$measures of health, $Z_3=$Measures of social unrest, $Z_4=$cost of ameliorating policies to be enacted..  The forms of the marginal utility functions then needed to be specified. For social unrest, health and education was assumed exponential, whilst the utility on cost was assumed linear.  It was therefore decided that one family of appropriate utility functions might take the form:
\begin{equation}
U(z)=a+bz_4+\sum_{i=1}^3 1-exp({-c_i z_i}),
\label{Eq:WarwickshireUtility}
\end{equation}
where $z=(z_1, z_2, z_3, z_4)$ and whose parameters $(a, b, c_1, c_2, c_3)$ were then elicited. As follows, observable variables are defined as proxies for the attributes required to compute the utility function in (\ref{Eq:WarwickshireUtility}).

\subsection{Measuring the attributes in the utility function}
\label{UF2}
The utility function depends on the proxy variables of health and education which are defined as follows.\\

\textbf{Health:} Suppose the expert panellists define a proxy as a function of number of admission to hospital with diagnosis of malnutrition (primary or secondary) and number of deaths with malnutrition listed on the death certificate either as primary or secondary cause. Admissions data are available in the Hospital Episode Statistics (HES) from the UK government's Health and a Social Care Information Service which routinely links UK Office for National Statistics (ONS) mortality data to HES data. 
In the UK, the number of deaths caused primarily by malnutrition are very low and rates are not significantly different over time. Besides, malnutrition is usually accompanied by other diagnoses such as diseases of digestive system, cancers, dementia and Alzheimer’s disease. Thus, the increase of deaths with malnutrition as a contributory factor might be due to ageing of the population and not due to food insecurity. Regarding admissions with malnutrition even the primary diagnosis numbers have increased over time with 391 in 2007-08 and 780 in 2017-18. Thus, in this work we considered the primary and secondary admission cases as a proxy for the health variable. Thus, the variable Health is defined as the count of finished admission episodes with a primary or secondary diagnosis of malnutrition coded ICD-10. A ICD-10 code of malnutrition on the episode indicates that the patient was diagnosed with, and would therefore being treated for malnutrition during the episode of care. \\

\textbf{Education:} The proxy for education could be defined as a function of educational attainment such as the proportion of pupils achieving expected grades in key stages 1, 2 and 4. Even though educational attainment is published annually at local and national levels by the UK government's Department for Education, the score system has changed in previous years and temporal comparisons are not adequate \citep{DeptEduc14}. Thus, as a proxy for education and its relation to food security we considered the proportion of pupils at the end of key stage 4 who were classified as disadvantaged. Thus, the variable Education is measured as the percentage of pupils at Key Stage 4 who were classified by the Department for Education as disadvantaged including pupils known to be eligible for free school meals (FSM) in any spring, autumn, summer, alternative provision or pupil referral unit census from year 6 to year 11 or are looked after children for at least one day or are adopted from care. Before 2015 this classification considered those who have been eligible for Free School Meals at any point in the last 6 years and Children who are ‘Looked After’. In 2015 this definition was widened to also include those children who have been ‘Adopted From Care’.  Pupils classified as disadvantaged have a lower average educational attainment record than other pupils and there is a direct correlation between level of qualification and unemployment in later life;  Poor educational attainment is strongly correlated with teenage pregnancy, offending behaviour, and alcohol and drug misuse. Comparisons between educational attainment for disadvantage and other pupils indicate a difference of 4.07 (2010/2011) and 3.66 (2016/2017) in the attainment gap index for Key stage 4 for state funded schools in England. The gap index are scores measuring the differences between the disadvantaged and non-disadvantaged groups in Key level 2 and 4 \citep{DeptEduc14}. The index is the mean rank for all the disadvantaged and non-disadvantaged pupils divided by the number of pupils in each cohort. This decimal mean rank difference is scaled to 10 and ranges from 0 to 10, where a higher value means a higher attainment of non-disadvantaged compared to disadvantaged pupils. The index aims to be resilient to changes in the grading systems and in the assessments and curricula, and may be used for temporal comparisons.\\

\textbf{Social Unrest:} Inadequate food security can cause food riots \citep{Lagi2012a}. In the UK, a riot is defined by section 1(1) of the Public Order Act 1986 as where 12 or more persons who are present together use or threaten unlawful violence for a common purpose and the conduct of them (taken together) is such as would cause a person of reasonable firmness present at the scene to fear for his personal safety, each of the persons using unlawful violence for the common purpose is guilty of riot. Riot data is collected by the police. Whilst the likelihood of a food riot is small in the UK currently, post-riot repairs both to physical environment and community relations can be considerable.
\\

\textbf{Costs} Costs of candidate intervention policies are routinely calculated and form part of the decision-making process.  Indeed, as a response to falling budgets, decision makers might revise the criteria for assistance of various kinds, for instance by making the eligible cohort smaller. Interventions which are effective but budget-neutral or cost-saving are obviously preferred, however, when the benefit of intervention may not be seen within the same financial year, this would form part of the decision-makers' discussion after the policies had been scored. This is the approach we take here, by scoring the policies and leaving the costs for final discussions of decision makers.
 
\subsection {Structure of the IDSS}

\begin{figure}[htb]
\centerline{\includegraphics[width=0.9\textwidth]{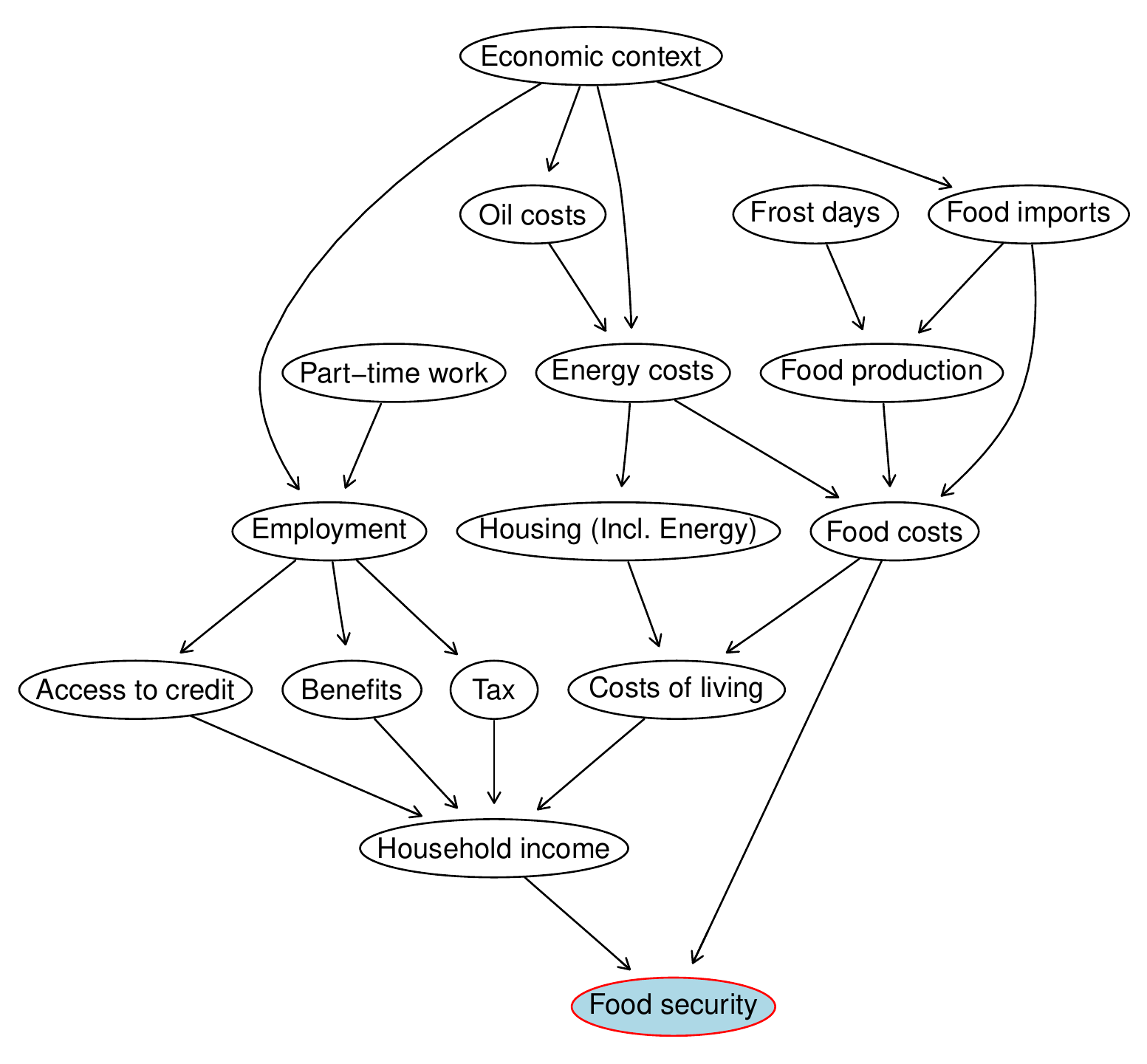}}
\caption{\footnotesize IDSS proposed for UK food security decision support. } 
\label{UKDBN}
\end{figure}

Having found a parsimonious form of utility function, we are able to begin to build the architecture of the supporting structural model. The paradigm we used for this is described in detail in \citep{Smith2010}.  The method involves first eliciting those variables which directly influence the attributes of the utility function, then the variables which affect those variables and so on until a suitable level of detail has been obtained. This was effected using an iterative process, drawing on the food poverty literature and checking with domain experts, refining and repeating.   In particular, the general framework was confirmed by work produced independently in \cite{LoopstraThesis2014}. The variables and their dependencies for the UK food system are shown in Figure \ref{UKDBN}.

There are a range models which can be used for the overarching model of an IDSS, as listed in \cite{Smith2016}, and for the purpose of the IDSS for food security we selected a dynamic Bayesian network (DBN) as summarised in subsection \ref{subBN}. The structure was assumed to be fixed over time.

Figure \ref{UKDBN} illustrates the  16-node DBN obtained through literature and confirmed by the experts. The node food security represents the two variables, health and education, considered in the utility function.

\newpage
\clearpage

\subsection{Expert panels}

Having identified the factors influencing household food security in the UK the next step is to identify the most relevant experts to provide information on these. The panels constituted for such an IDSS will often be chosen to mirror the panels that are already constituted for similar purposes, e.g. in the UK, the Office for Budget Responsibility, HM Treasury and The Confederation of British Industry all produce economic forecasts on the UK Economy. Looking at where the relevant information is held gives some very natural panels. 

\begin{figure}[htb]
\centerline{\includegraphics[angle=270,width=0.9\textwidth]{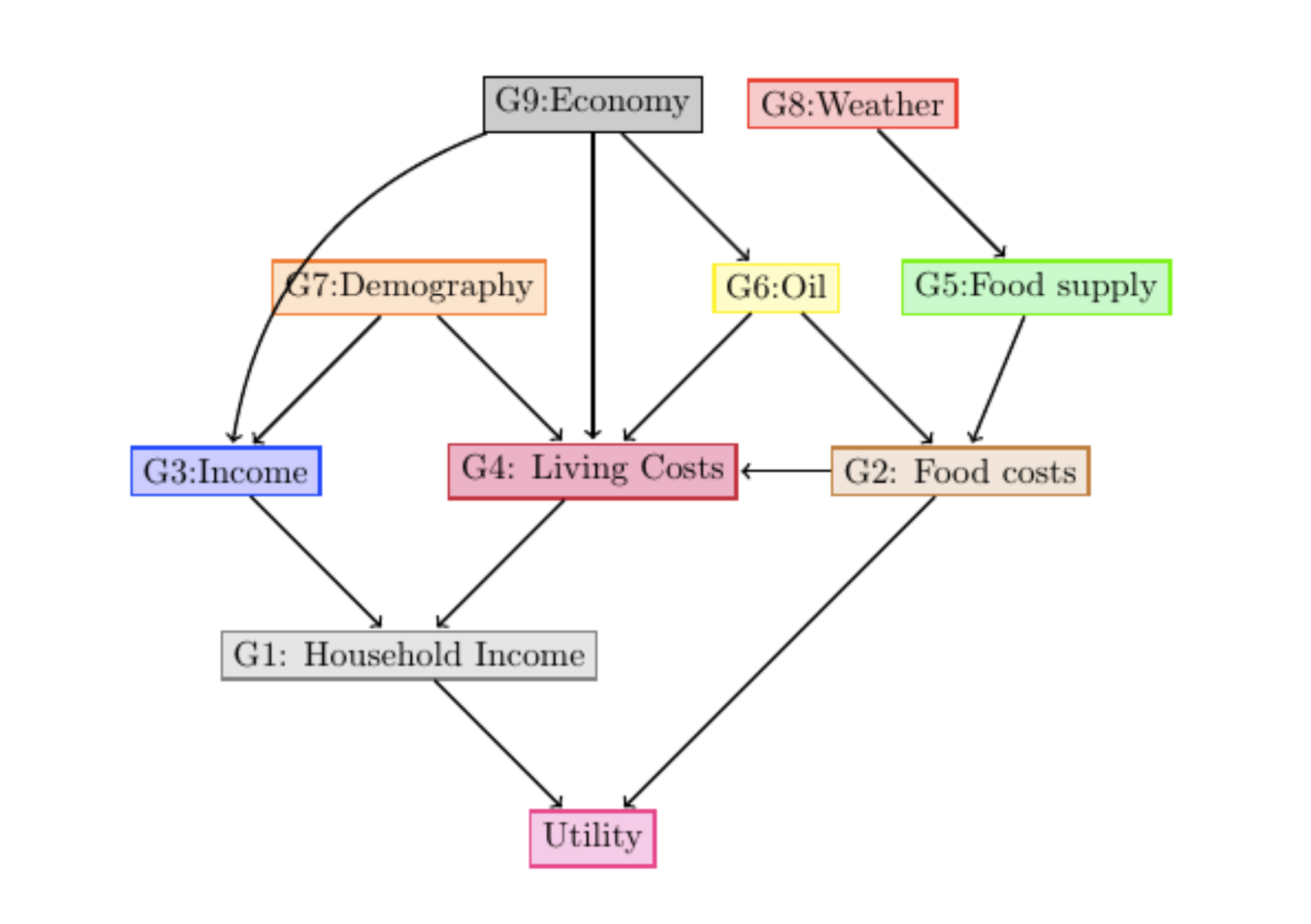}}
\caption{The expert panels required for this IDSS. Each node represents an expert panel which, using its models and data, provides summaries of expected values and relevant moments under each policy decision being considered.}
\label{FoodDBNTikz}
\end{figure}

The 16-node DBN illustrated in figure \ref{UKDBN} becomes a 9-panel IDSS (figure \ref{FoodDBNTikz}). Panel G2 reports on cost of food given inputs from pane G5 on food supply, incorporating imports and exports, domestic food production and supply chain disruption. Panel G5, in turn, relies on information from G8 the Met office on weather and climate patterns to calculate its expectations of food supply, since both domestic and world production and supply chain disruption are weather related. Household income, G1, impacts directly on the utility. Panel G1 relies on information provided by G3 and G4 to make its predictions under different policy scenarios. G4 adivises on cost of living including energy, housing and other essentials. G3 assesses income taking into account employment, tax and social security, taking inputs from G7 and G9. G7 advised on demography, including single parents, immigrants, disability and those with no recourse to public funds. G9 advises on matters of the economy 
and informs the oil price panel, G6, and the cost of living panel, G4 as well as G3.

\subsection{Dynamical Bayesian Network IDSS for food security}

Here we assume plausible models for the expert panels and utility, based on publicly available data.

The attributes being measured to compose the food network were obtained at the Office for National Statistics which publishes official statistics for the UK. The time series for all nodes are measured yearly and the temporal window considered goes from 2008 to 2018. Each variable is detailed at Appendix A. 

\vspace{0.3cm}
For the purposes of this proof of concept, social unrest was omitted since there was no available data.
The health and education indicators are the attributes in the utility function and are directly affected by household income (HIncome, panel $G_1$) and food costs (CFood, panel $G_2$). The variables are modelled in the log scale as both are percentages or rates.
\begin{eqnarray}\nonumber 
log(Health_t) & = & \delta_{01,t}+\delta_{11,t} HIncome_t+\delta_{21,t} CFood_t +\epsilon_{ht},\\ \nonumber
log(Education_t) & = & \delta_{02,t}+\delta_{12,t} HIncome_t+\delta_{22,t} CFood_t +\epsilon_{et}.
\end{eqnarray}

Panel $G_1$ advises on household income aiming to reflect the amount of money that households have available after accounting for the expendures with living (panel $G_4$), taxes and also the access to credit and benefits (panel $G_3$).
\begin{eqnarray*} 
 HIncome_t & = & \theta_{01,t}+\theta_{11,t} Lending_t +\theta_{21,t} Tax_t+\theta_{31,t} Benefits_t+\theta_{41,t} CLiving_t+\epsilon_{1t}.
 \end{eqnarray*}

The variable costs of food (Panel $G_2$) depends on costs of energy (panel $G_6$) and on food supply, imports and exports and food production (panel $G_5$).
\begin{eqnarray}\nonumber 
 CFood_t & = &  \theta_{02,t}+\theta_{12,t}, FProduction_t +\theta_{22,t} FImports_t +\theta_{32,t} CEnergy_t+ \epsilon_{2t}.
\end{eqnarray}

Panel $G_3$ reports on variables affecting the income such as lending, tax and unemployment. Unemployment depends on the economic context (panel $G_9$) represented by GDP and on part-time workers (panel $G_7$).
\begin{eqnarray}\nonumber 
 Lending_t & = & \theta_{03,t}+\theta_{13,t} Unemployment_t +\epsilon_{3t},\\ \nonumber
 Tax_t & = & \theta_{03,t}^*+\theta_{13,t}^* Unemployment_t +\epsilon_{3t}^*,\\ \nonumber
 Benefits_t & = & \theta_{03,t}^{**}+\theta_{13,t}^{**} Unemployment_t +\epsilon_{3t}^{**},\\ \nonumber Unemployment_t & = & \theta_{03,t}^{***}+\theta_{13,t}^{***} \mbox{\it Part-time}_t + \theta_{23,t}^{***}GDP_t +\epsilon_{3t}^{***}.
\end{eqnarray}

Panel $G_4$ reports on costs of living which depend on costs of food (panel $G_2$), on costs of housing including energy. Costs of housing depends on costs of energy (panel $G_6$).
\begin{eqnarray}\nonumber 
 Cliving_t & = & \theta_{04,t}+\theta_{14,t} CFood_t+\theta_{24,t} CHousing_t + \epsilon_{4t},\\ \nonumber
 CHousing_t & = & \theta_{04,t}^*+ \theta_{14,t}^* CEnergy_t+ \epsilon_{4t}^*.
\end{eqnarray}
Panel $G_5$ (Food supply) reports on food production and imports  which depend on the economic context (panel $G_9$):
\begin{eqnarray}\nonumber 
 FProduction & = & \theta_{05,t}+\theta_{15,t} Gdp_t +\theta_{25,t} Imports_t  + \epsilon_{5t},\\ \nonumber
FImports_t & = & \theta_{05,t}^{*}+\theta_{25,t}^* GDP_t + \epsilon_{5t}^{*}.
\end{eqnarray}

Panel $G_6$ reports on oil costs and energy given inputs from panel $G_9$ about economic context.
\begin{eqnarray}\nonumber 
 COil_t & = & \theta_{06,t}+\theta_{16,t} GDP_t + \epsilon_{6t},\\ \nonumber
 CEnergy_t & = & \theta_{05,t}^{*}+\theta_{15,t}^{*} COil_t + \epsilon_{5t}^{*}.
\end{eqnarray}

Panel $G_7$ (Demography), $G_8$ (Weather) and $G_9$ (Economy) reports on demography, weather and economic context, respectively with model equations given by
\begin{eqnarray*}
log(PartTime_t) & = & \theta_{07,t},+ \epsilon_{7t},\\     
Frost_t & = & \theta_{08,t}+ \epsilon_{8t},\\
  Gdp_t & = & \theta_{09,t}+ \epsilon_{9t}.  
\end{eqnarray*}

Using these models as the panels' models, we now examine what happens to the utility under an number of scenarios.

\section{Model outputs and scenario evaluation\label{Out}}

\begin{figure}[htb]
\begin{center}
\begin{tabular}{ccc}
\includegraphics[width=3.8cm,height=3.2cm]{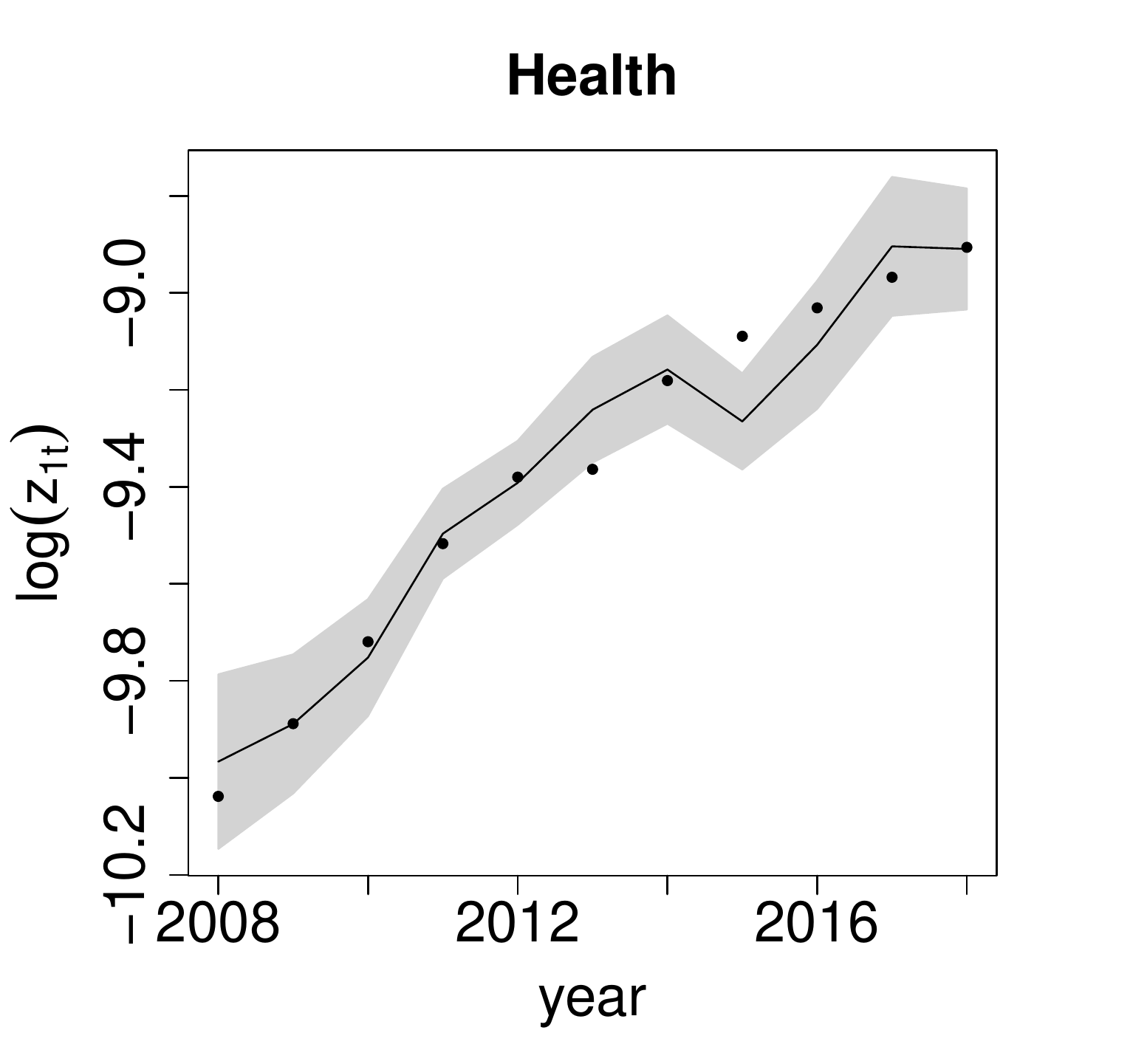} &
\includegraphics[width=3.8cm,height=3.2cm]{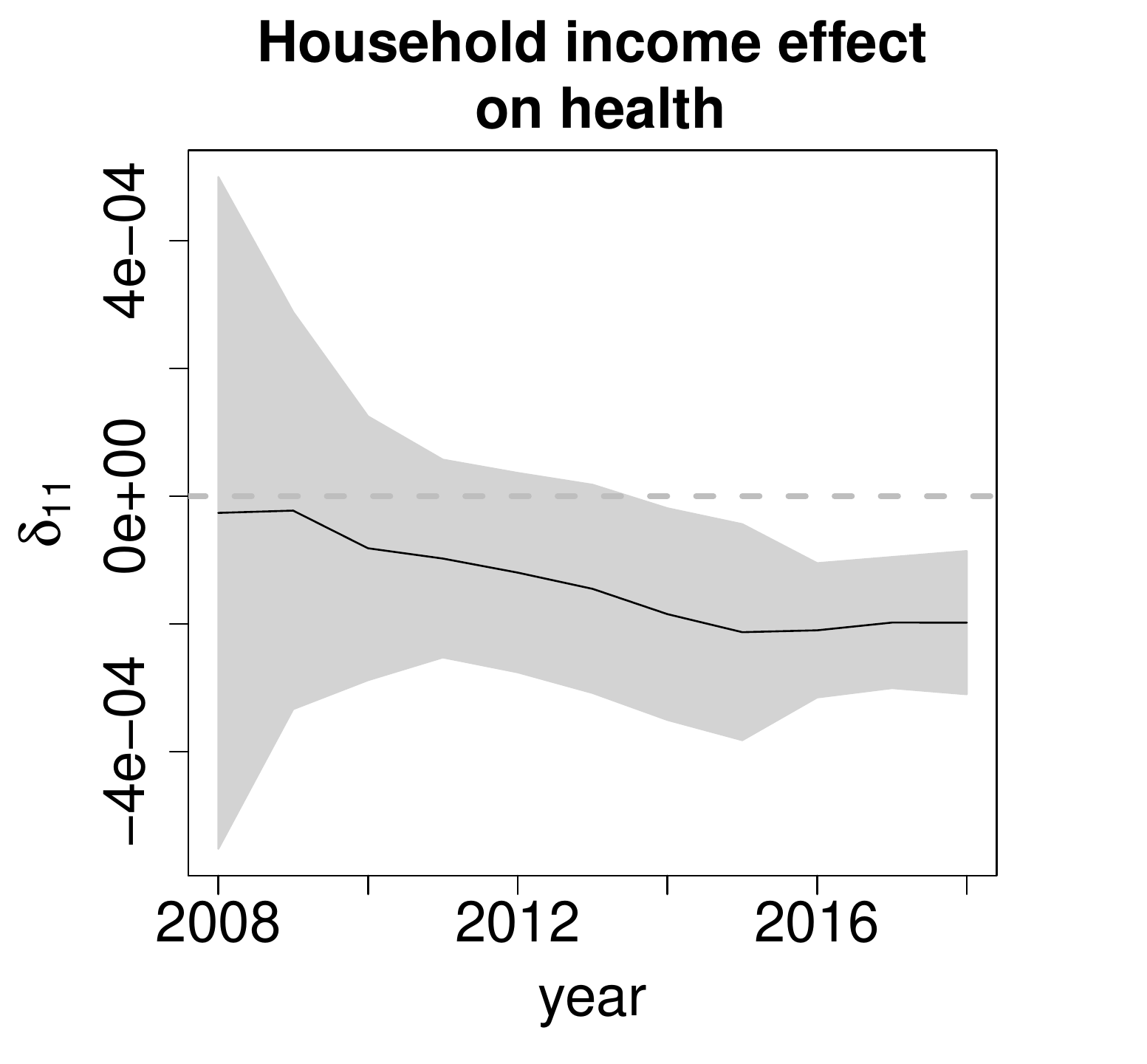} &
\includegraphics[width=3.8cm,height=3.2cm]{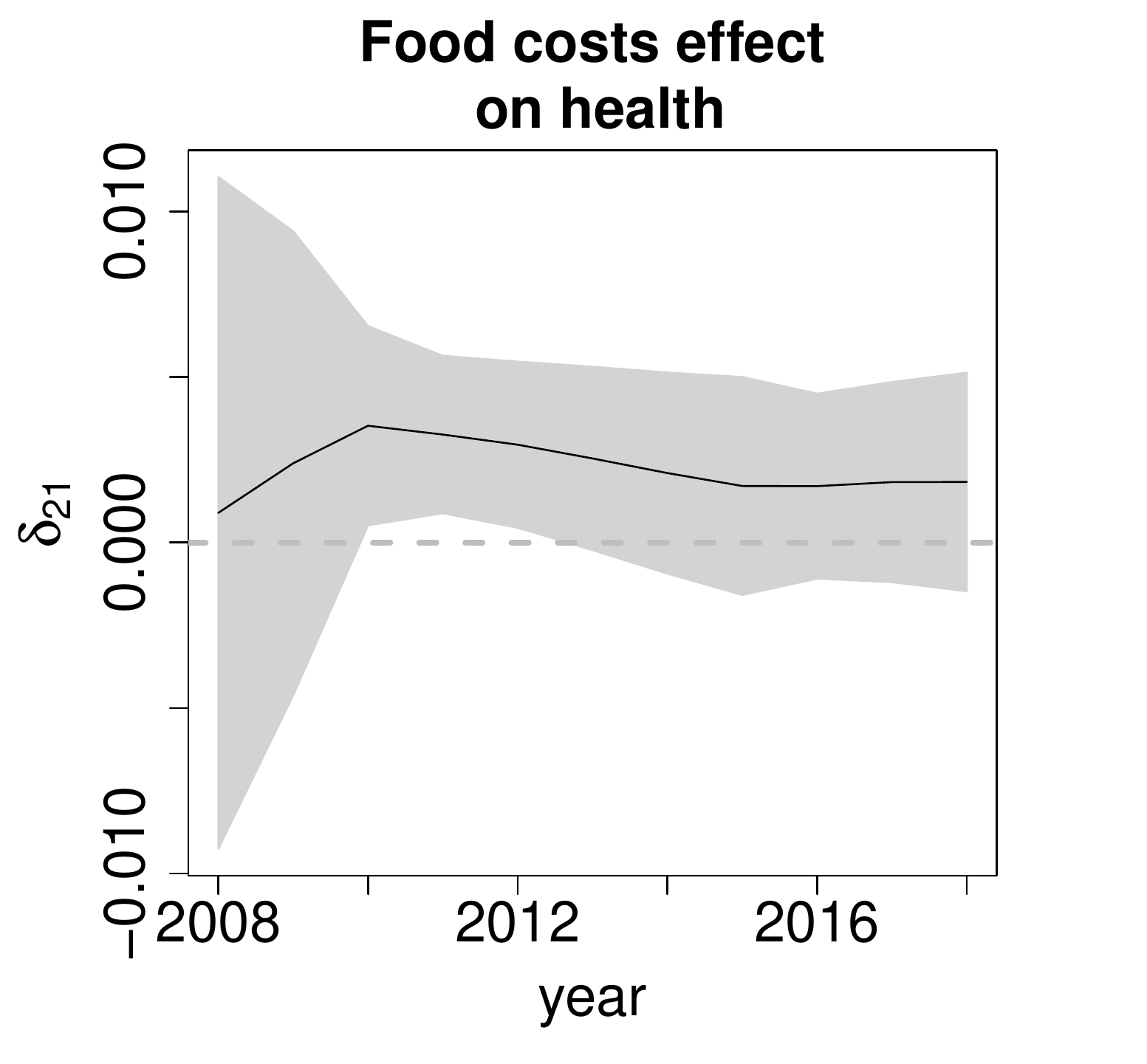} \\
\includegraphics[width=3.8cm,height=3.2cm]{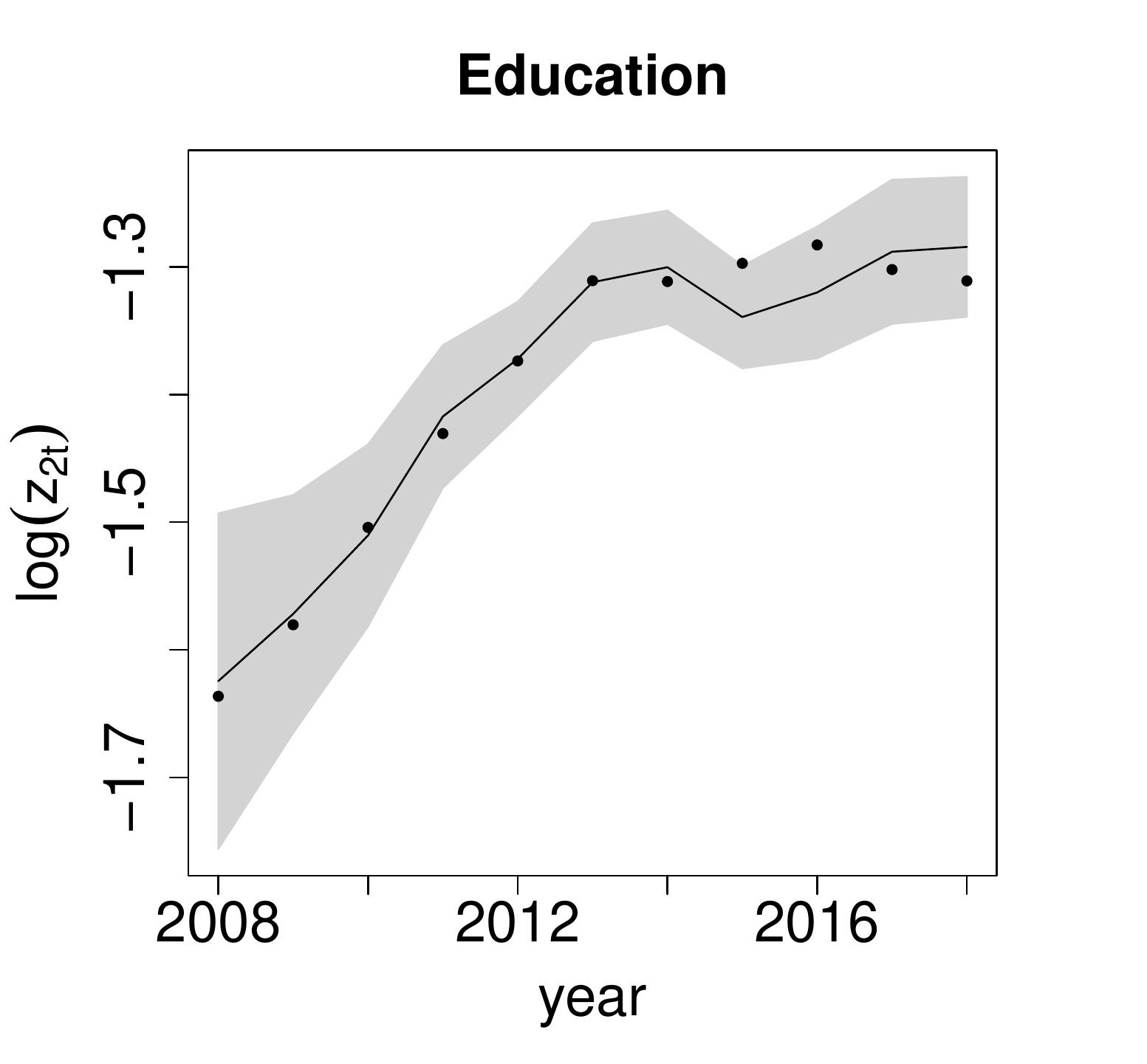} &
\includegraphics[width=3.8cm,height=3.2cm]{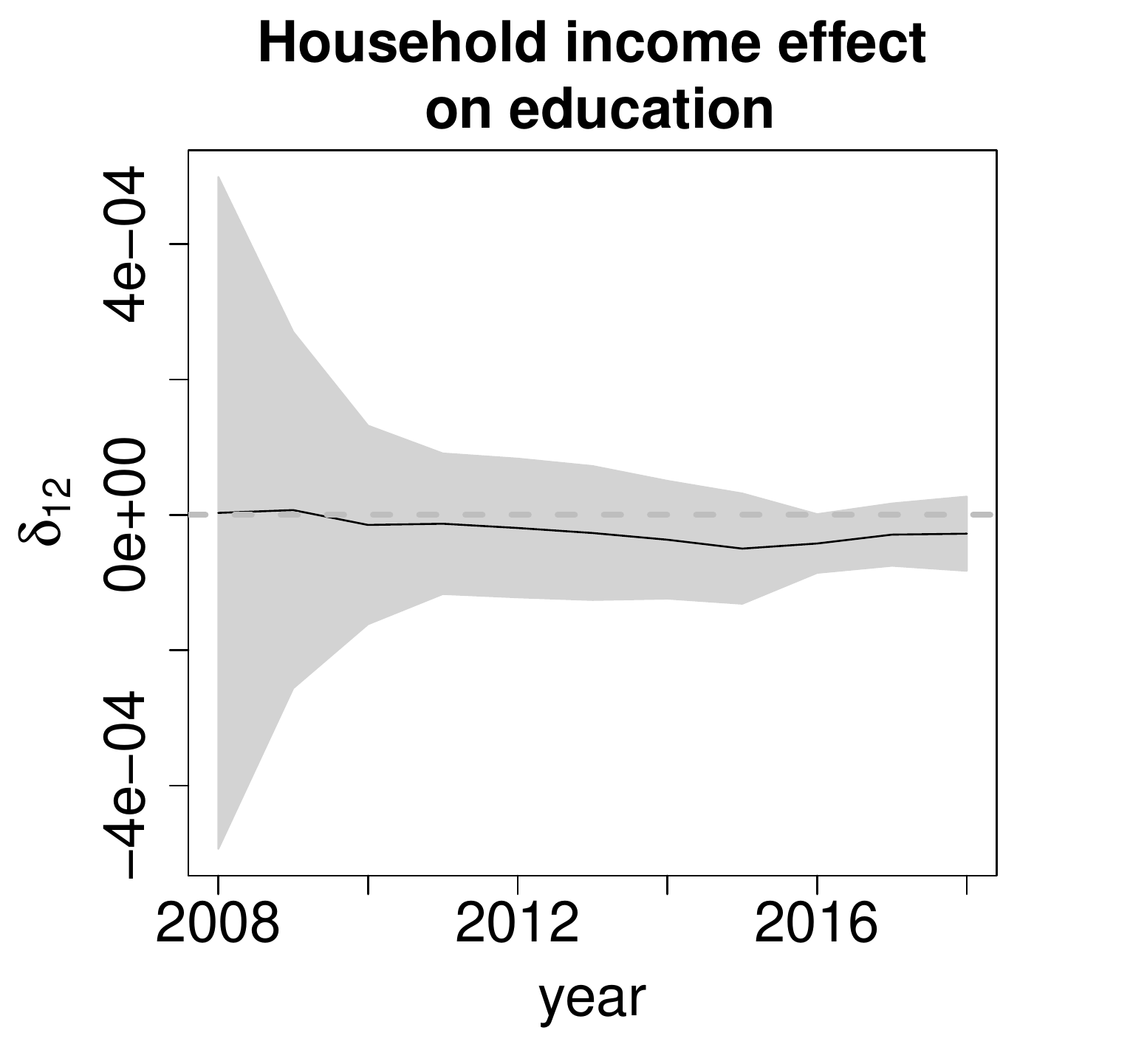} &
\includegraphics[width=3.8cm,height=3.2cm]{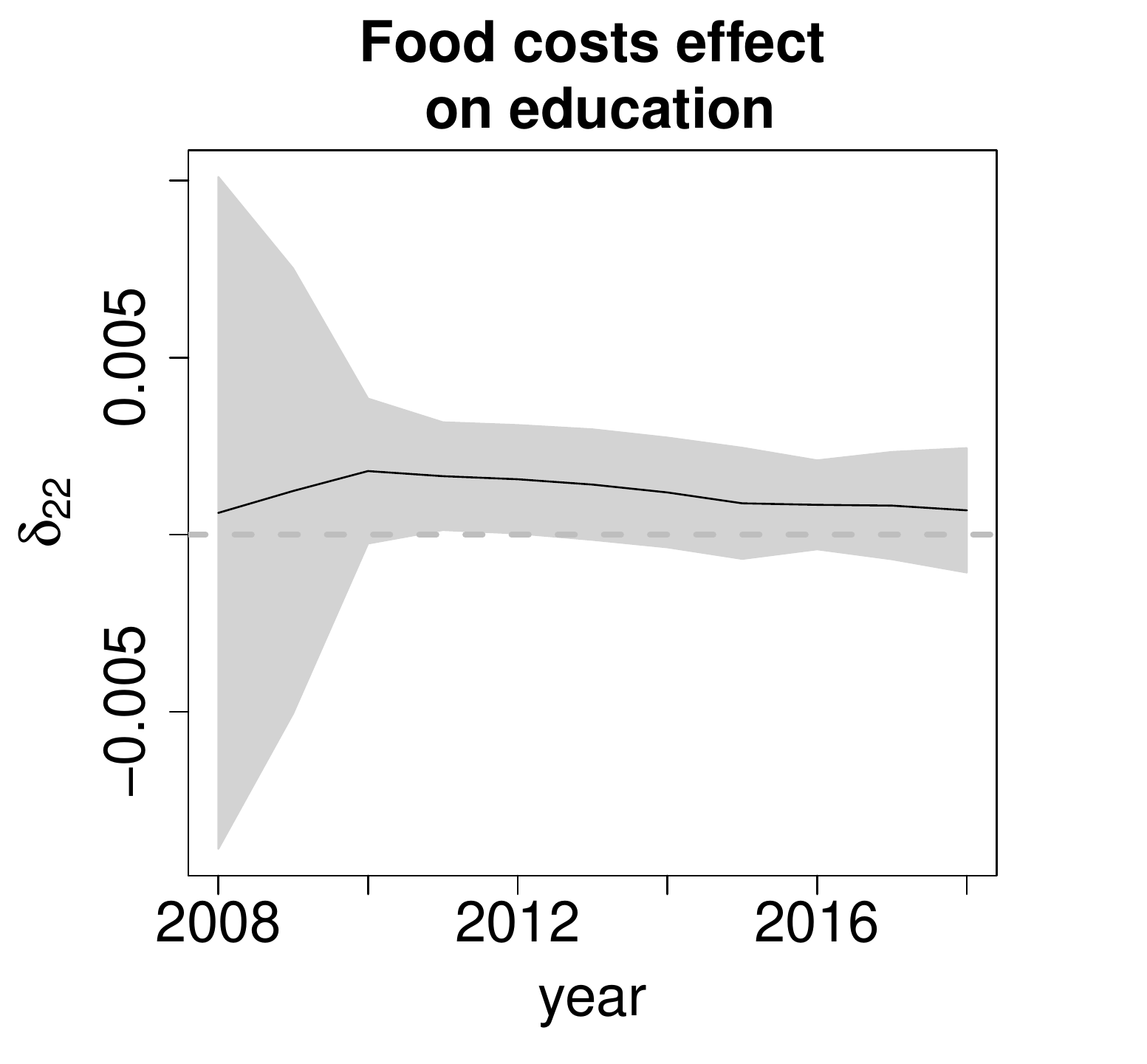} \\
\end{tabular}
\caption{Attributes composing the utility function, effects of household income and food costs and MDM fit (mean and $95\%$ credible interval), 2008-2018.}\label{Fig:fitz}
\end{center}
\end{figure}

\begin{figure}[htb]
\begin{center}
\begin{tabular}{ccc}
\includegraphics[width=3.8cm,height=3.2cm]{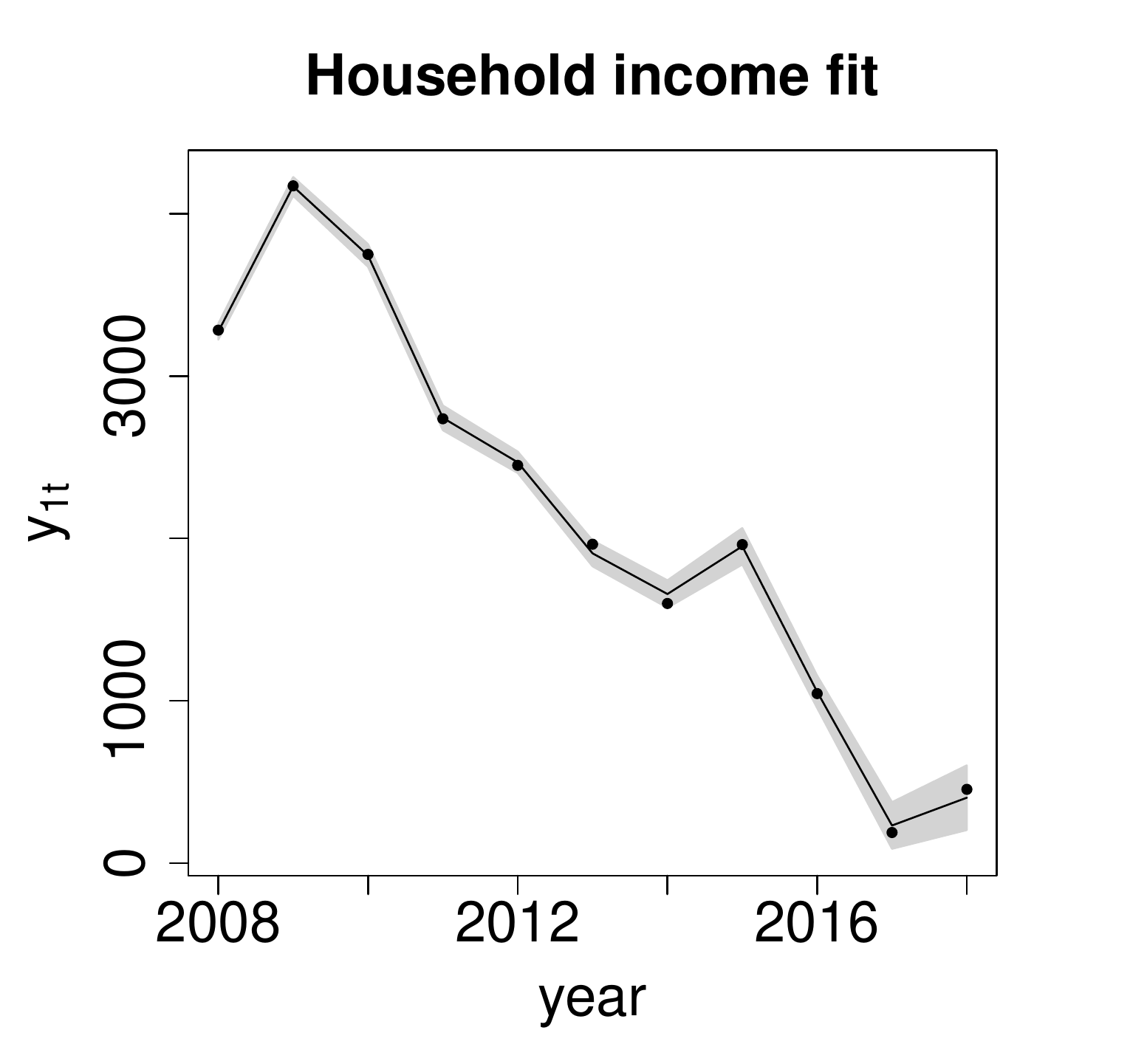} &
\includegraphics[width=3.8cm,height=3.2cm]{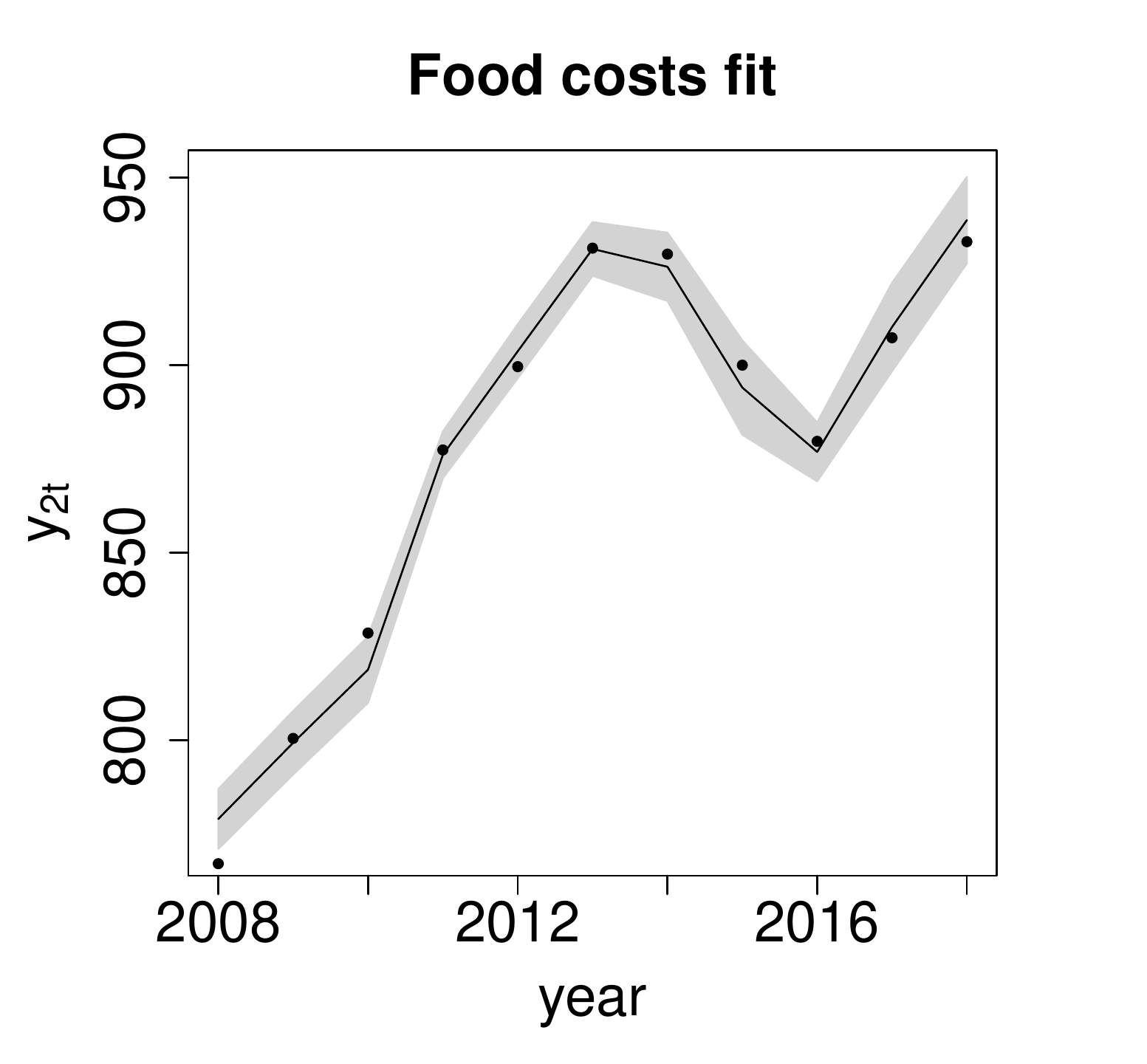} &
\includegraphics[width=3.8cm,height=3.2cm]{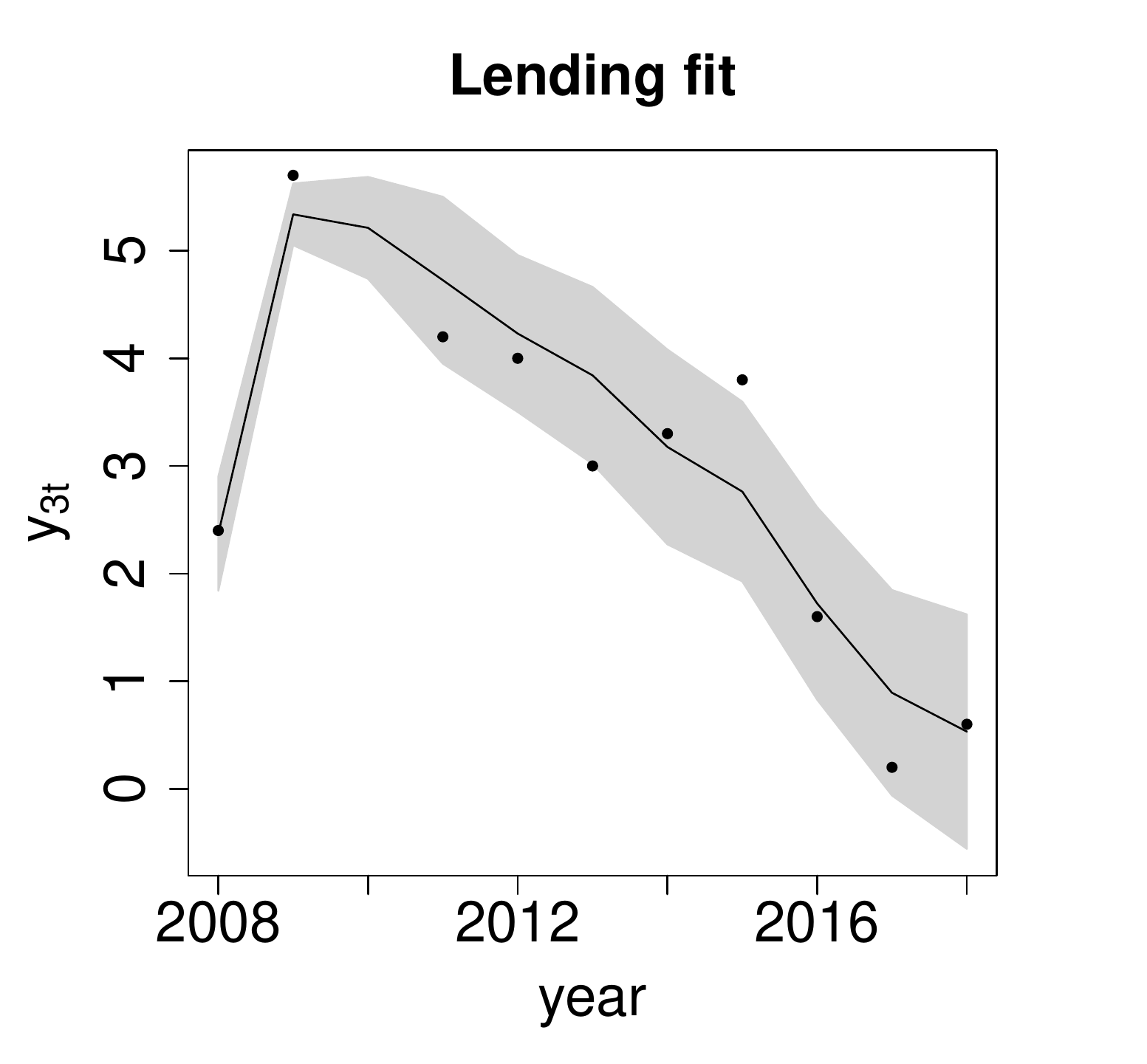}\\
\includegraphics[width=3.8cm,height=3.2cm]{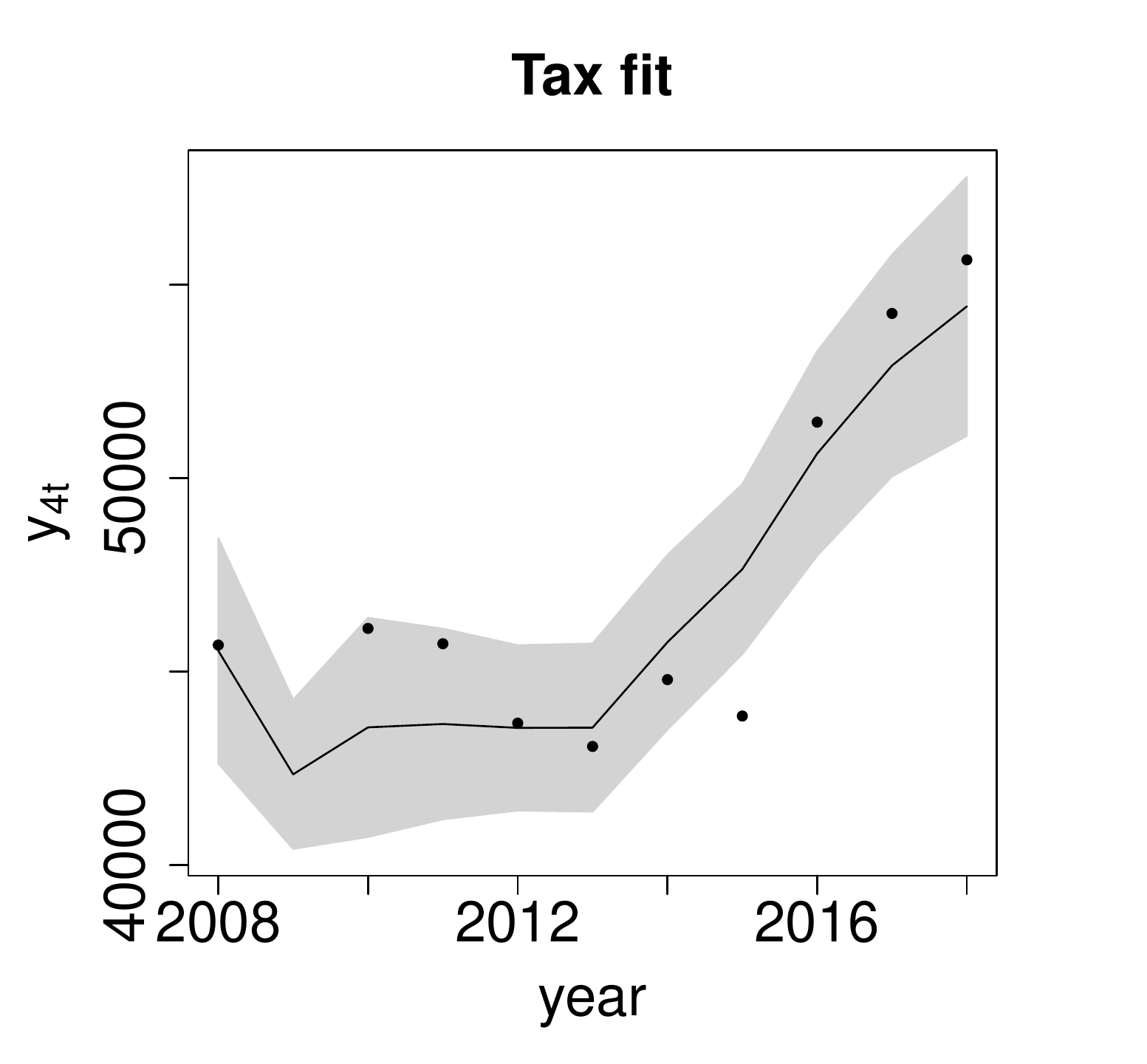} &
\includegraphics[width=3.8cm,height=3.2cm]{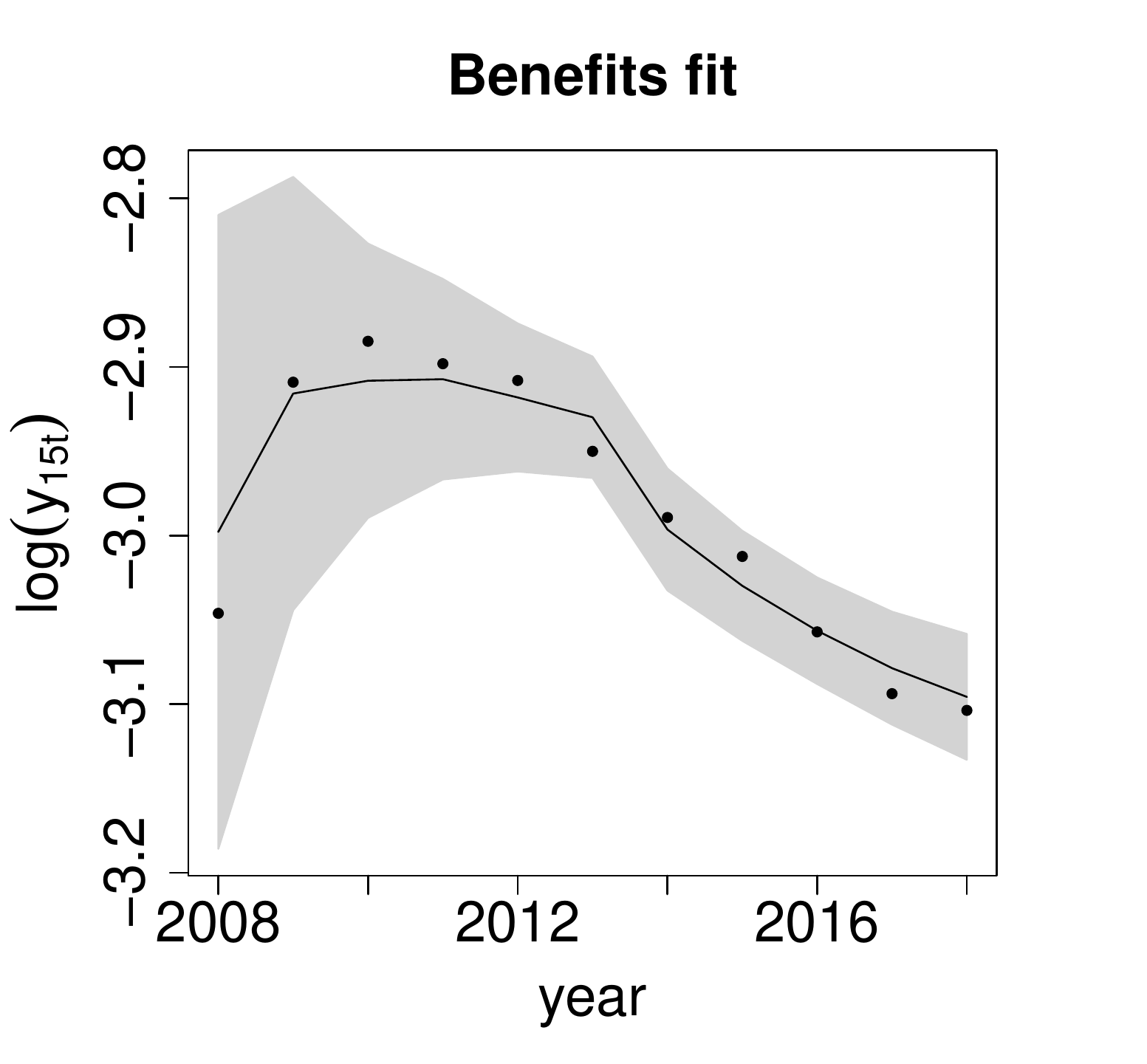} &
\includegraphics[width=3.8cm,height=3.2cm]{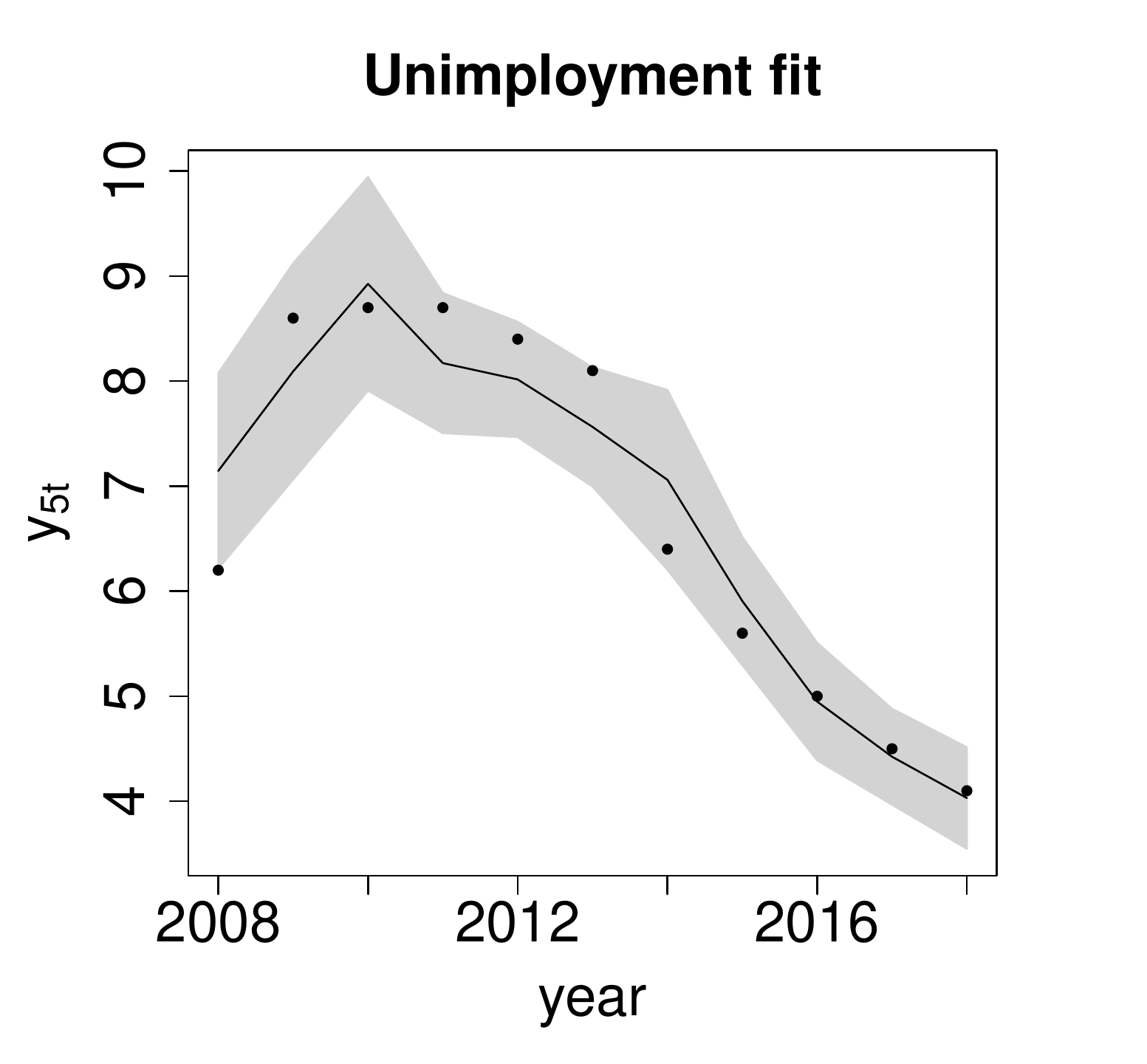} \\
\includegraphics[width=3.8cm,height=3.2cm]{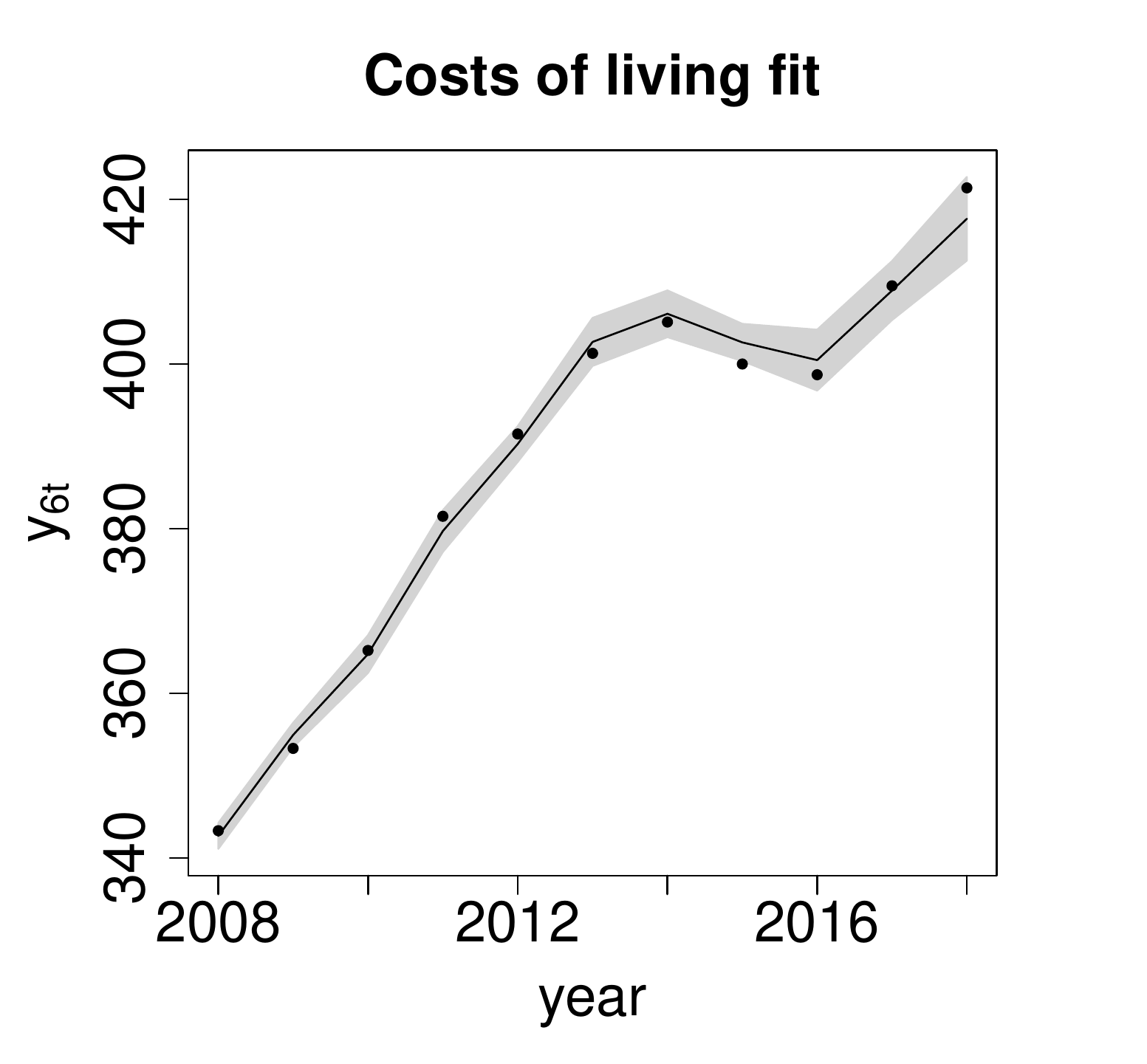}&
\includegraphics[width=3.8cm,height=3.2cm]{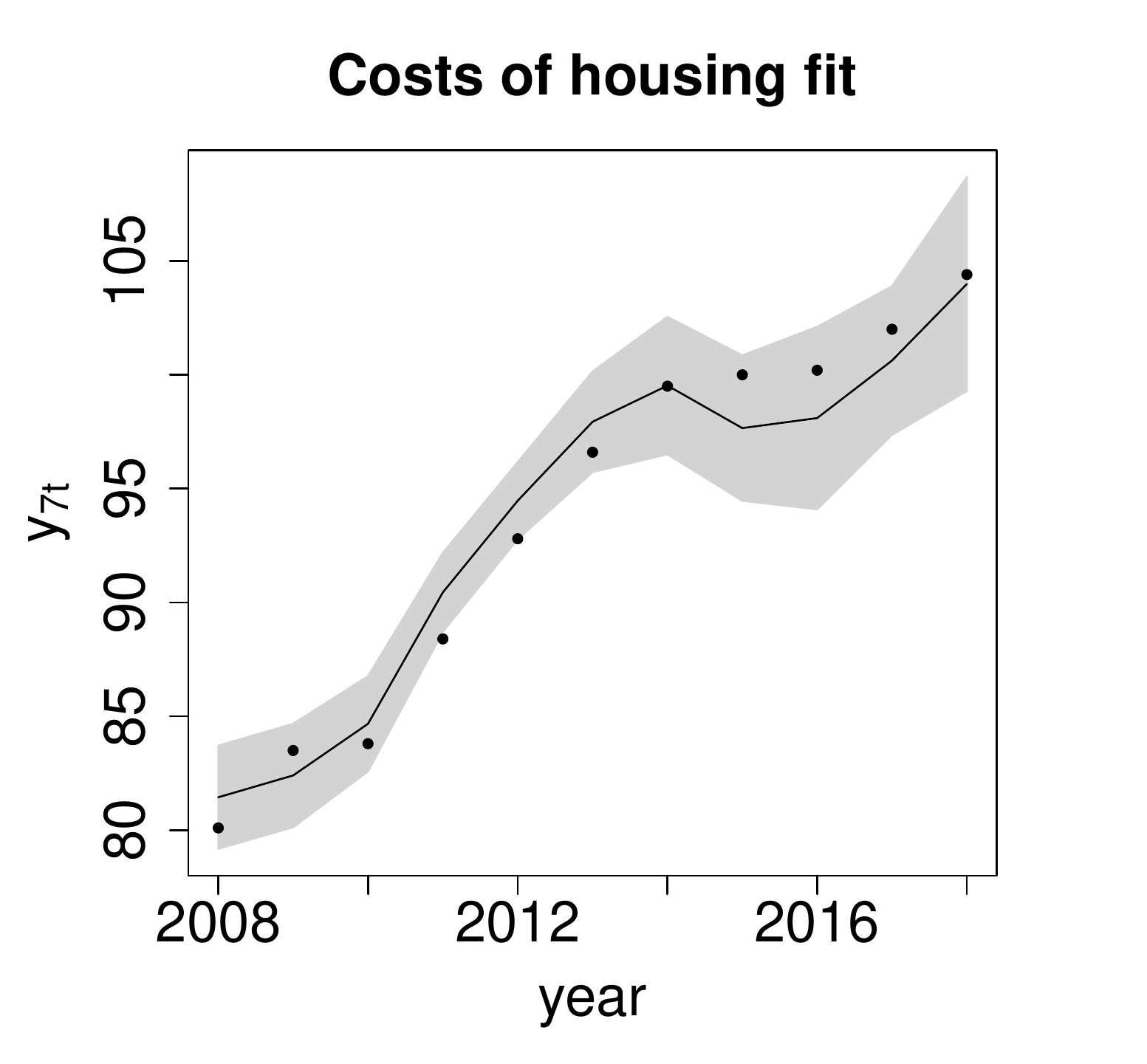} &
\includegraphics[width=3.8cm,height=3.2cm]{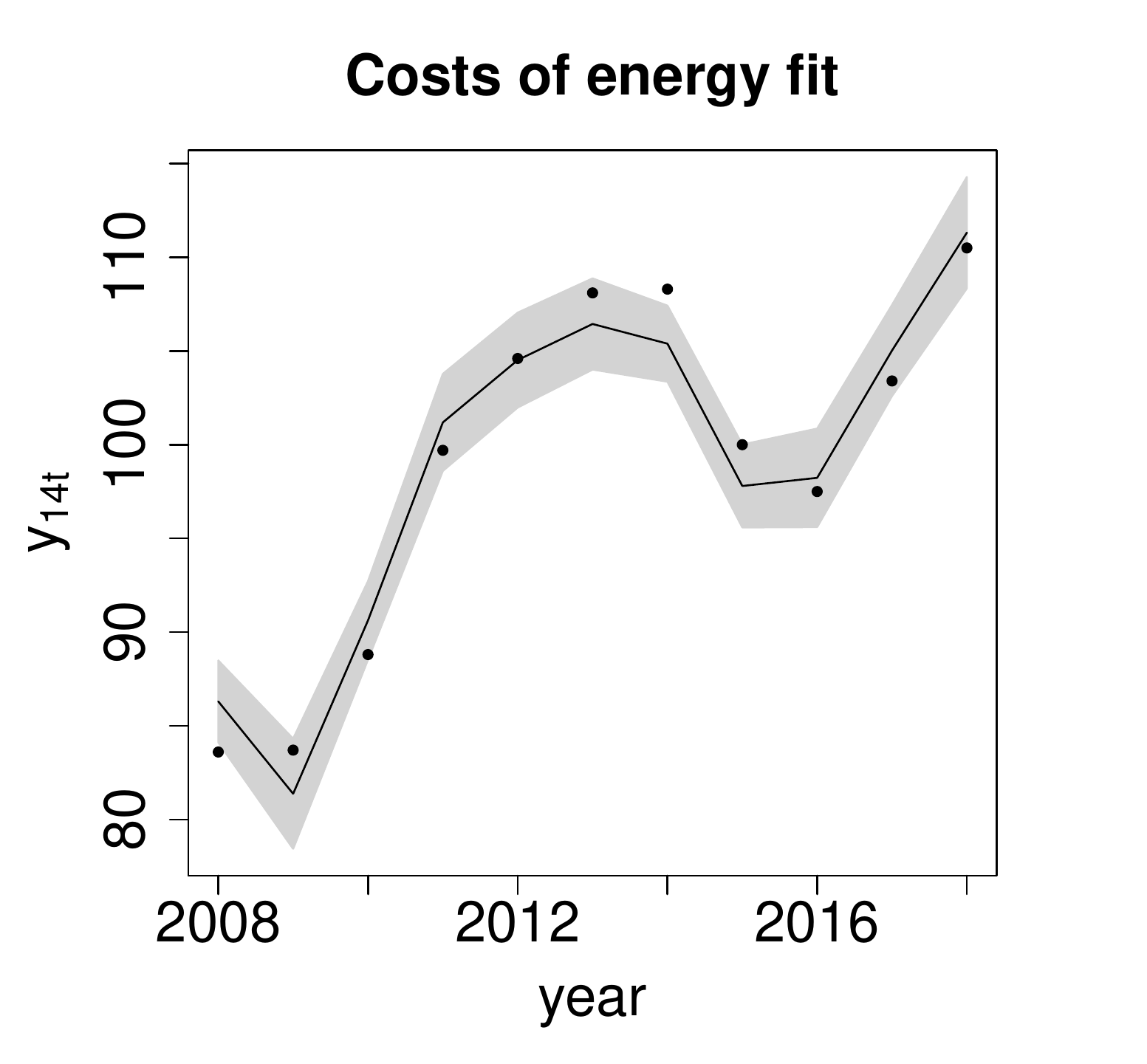} \\
\includegraphics[width=3.8cm,height=3.2cm]{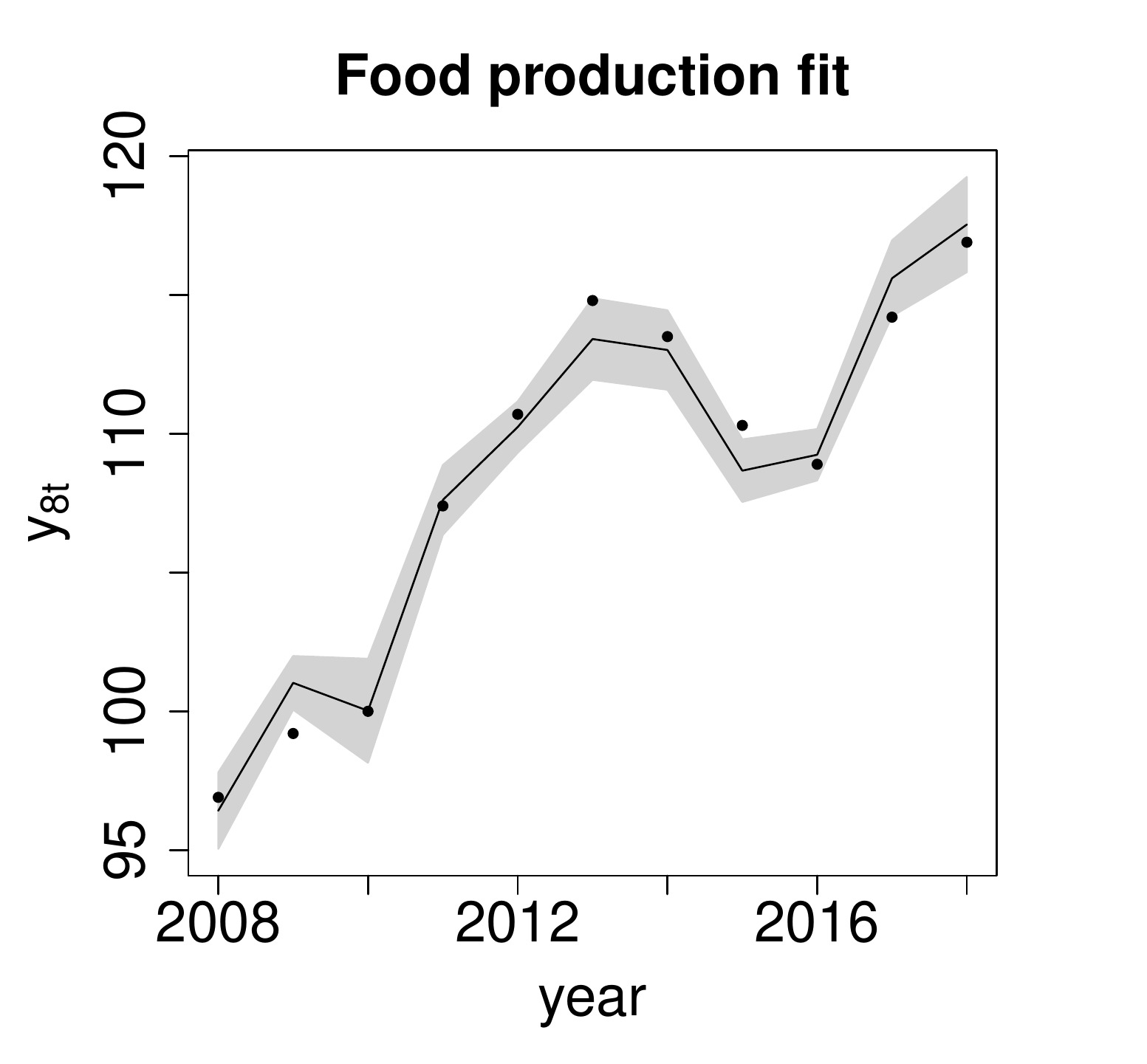}&
\includegraphics[width=3.8cm,height=3.2cm]{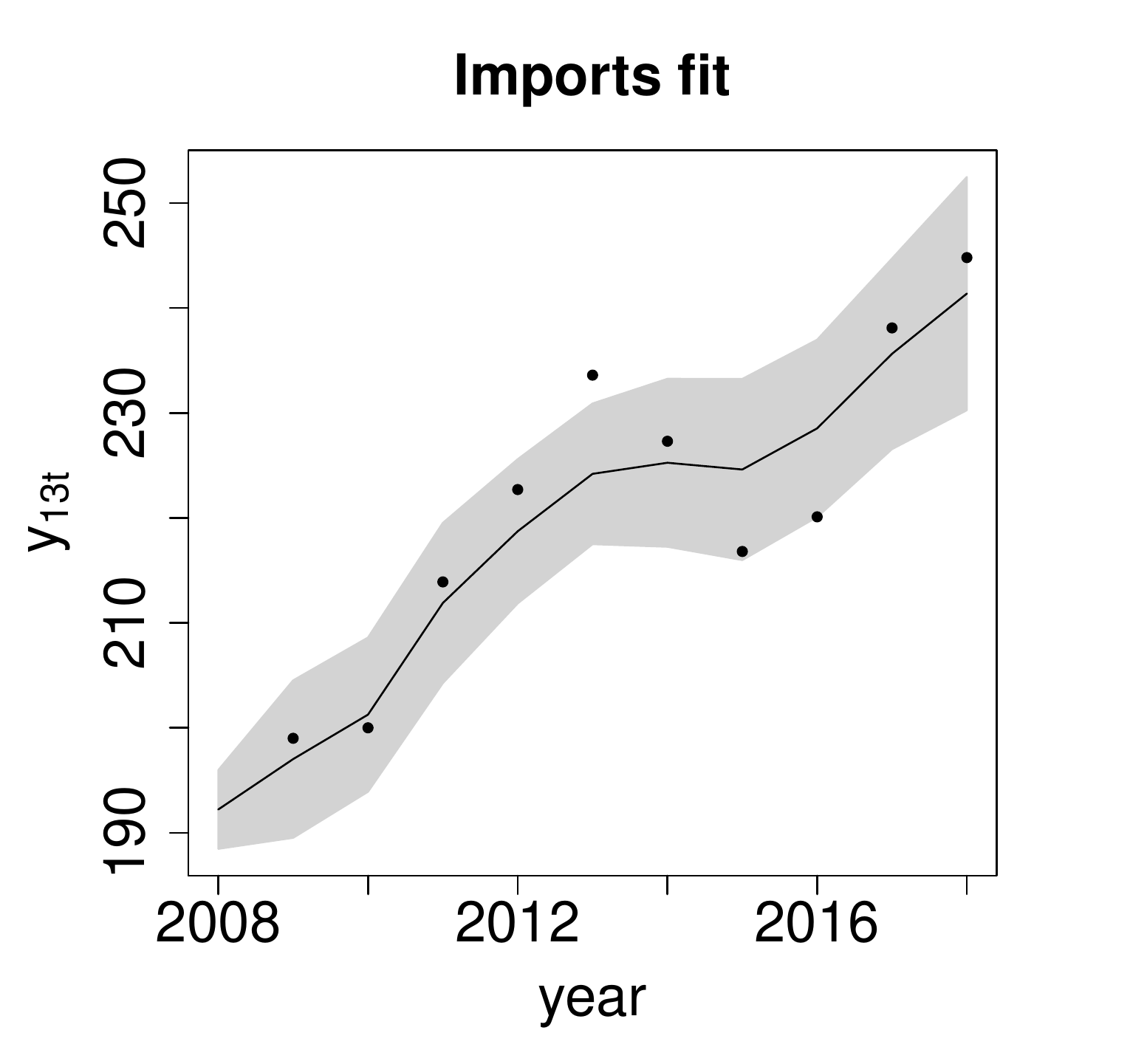} &
\includegraphics[width=3.8cm,height=3.2cm]{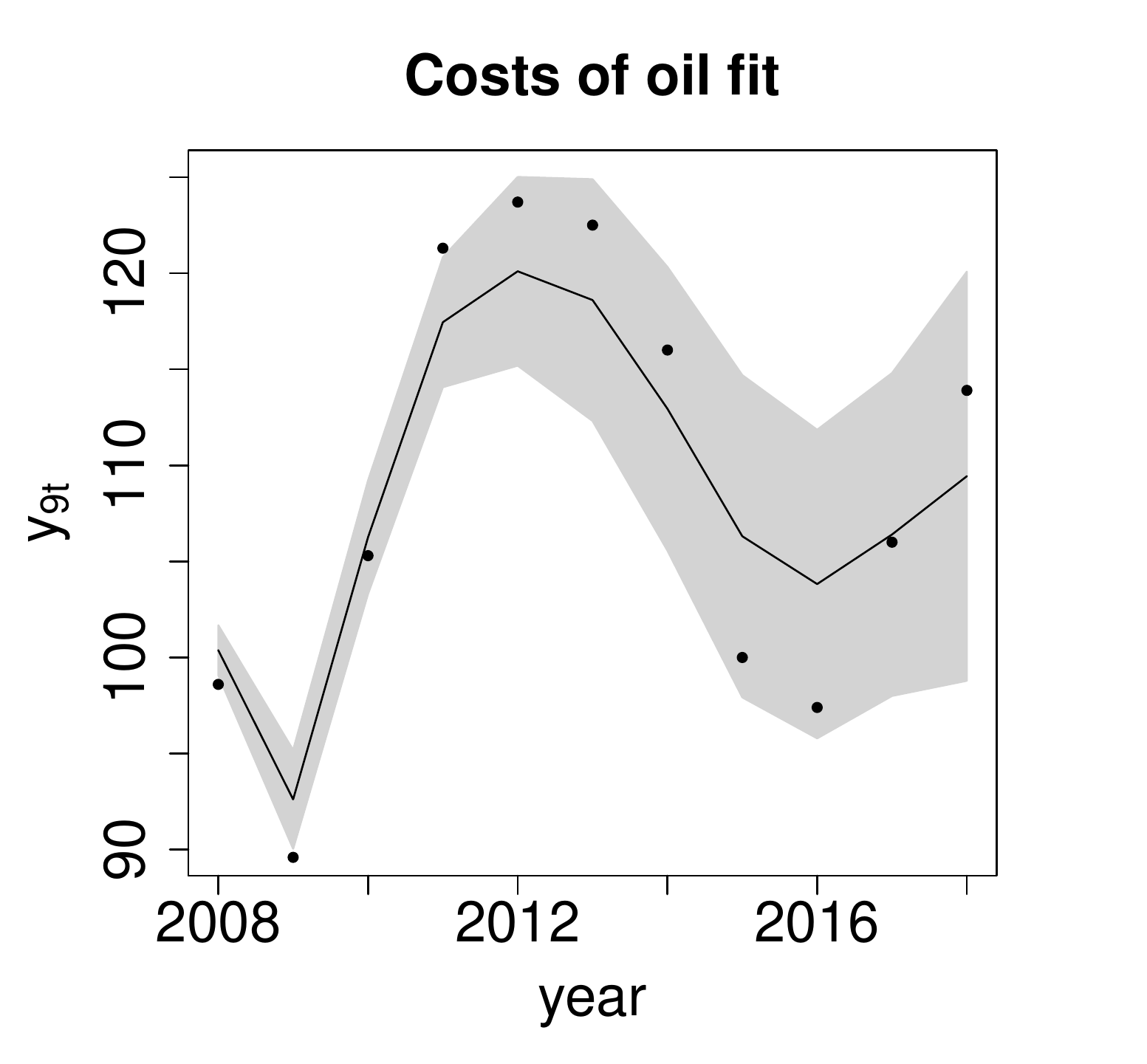} \\
\includegraphics[width=3.8cm,height=3.2cm]{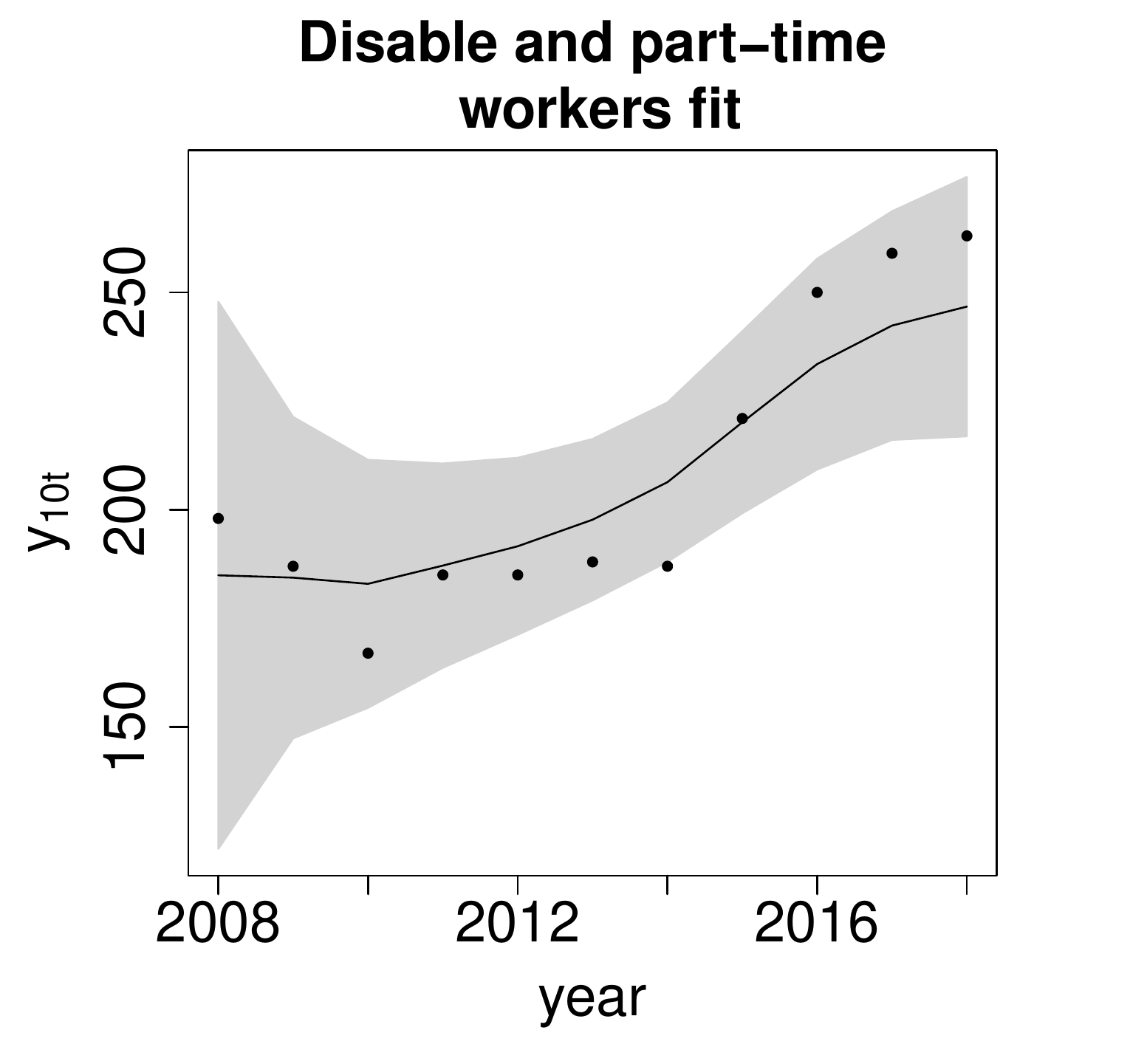}&
\includegraphics[width=3.8cm,height=3.2cm]{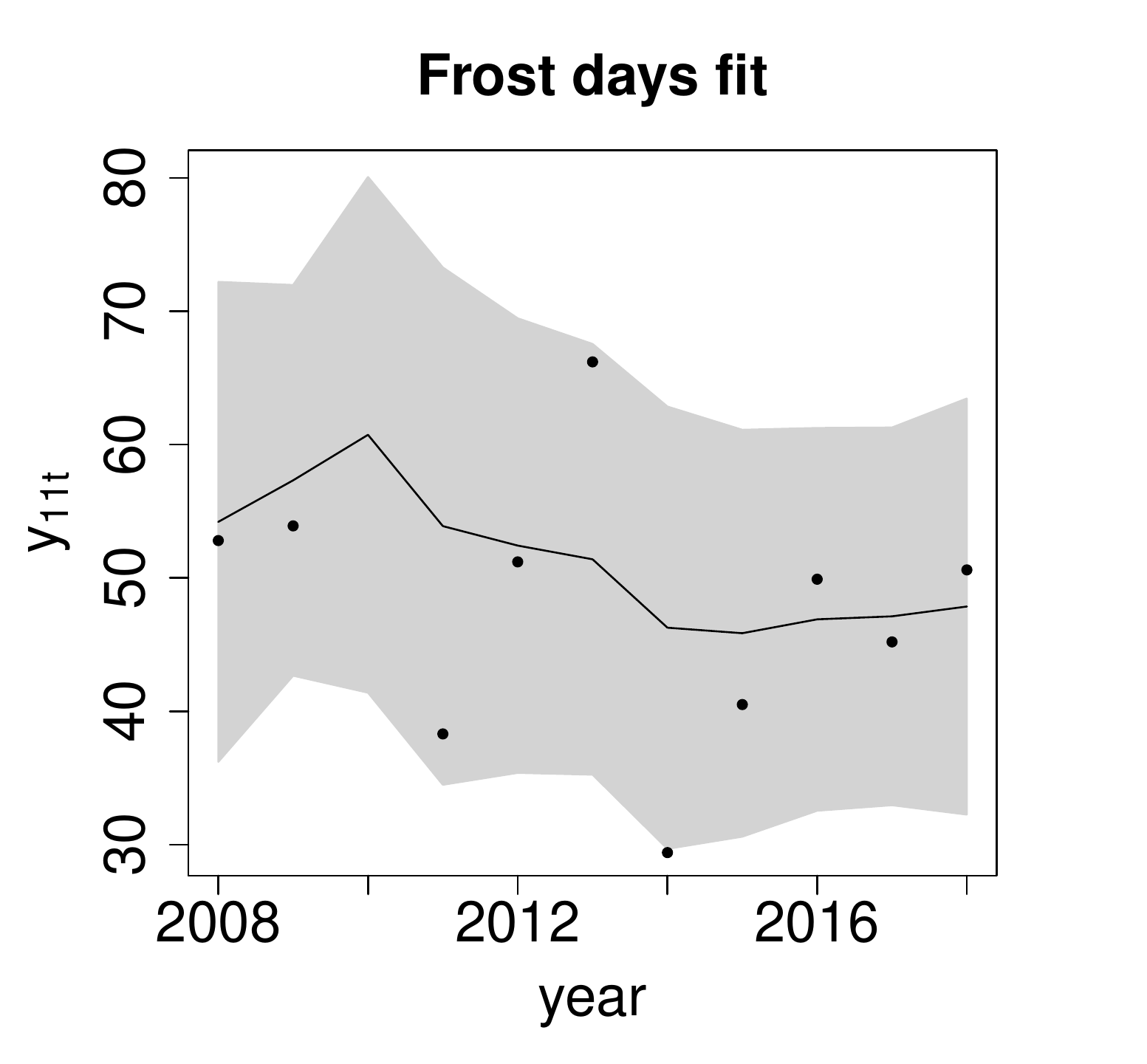} &
\includegraphics[width=3.8cm,height=3.2cm]{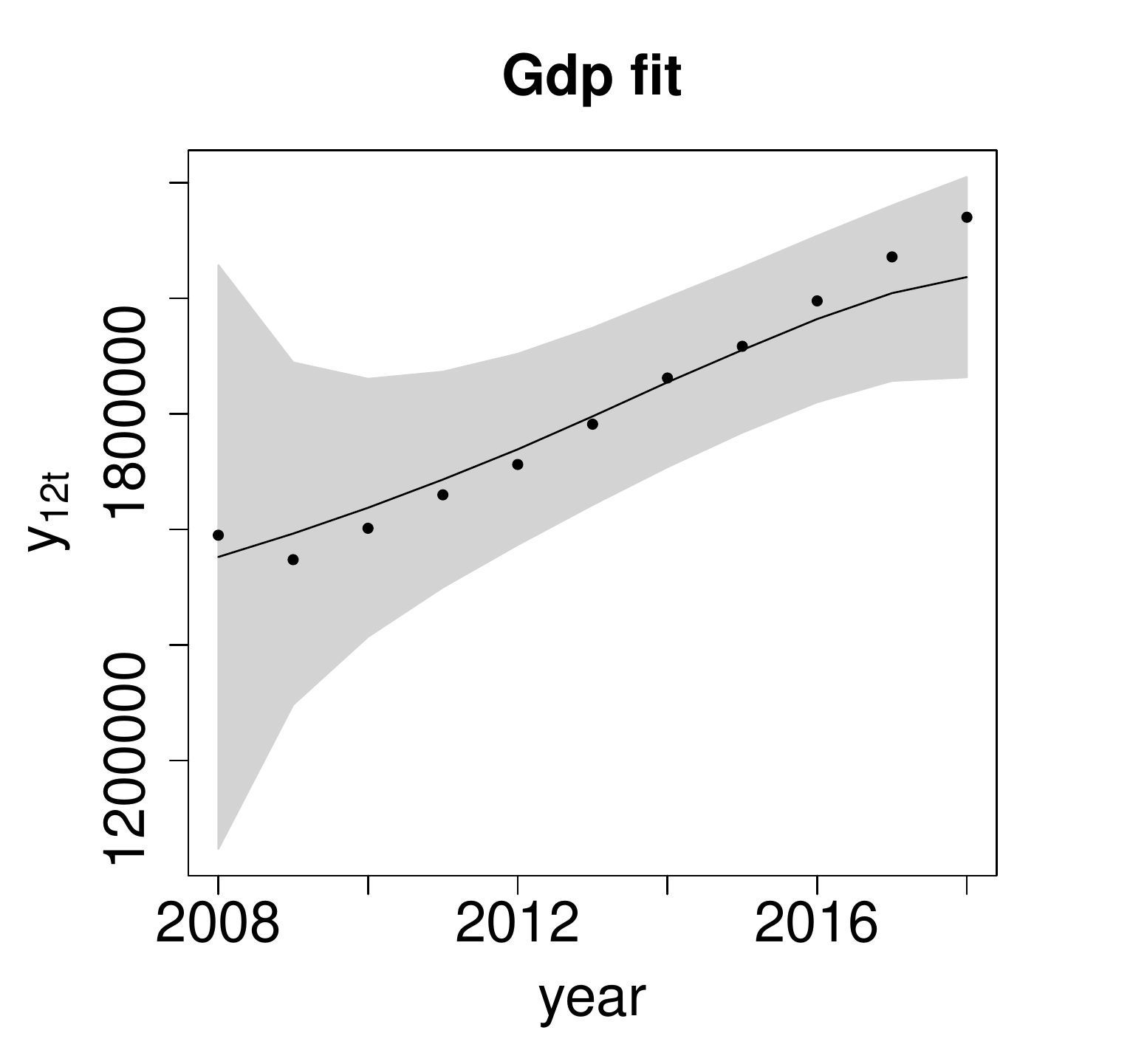}  
\end{tabular}
\caption{Variables composing the food network and dynamical regression model fit (mean and $95\%$ credible interval), 2008-2018.}\label{Fig:fity}
\end{center}
\end{figure}

Figure \ref{Fig:fitz} presents the fit and effects of household income and food costs on health and education obtained by recursively updating of posterior moments based on the forward filtering and backward algorithm presented in subsection \ref{subBN}. Notice the negative effect of household income and positive effect of food costs on the rate of malnutrition and percentage of disadvantaged pupils. Figure \ref{Fig:fity} presents the fit for all the variables in the food security network. 

\begin{figure}[htb]
\begin{center}
\begin{tabular}{cc}
\includegraphics[width=5.8cm,height=5.2cm]{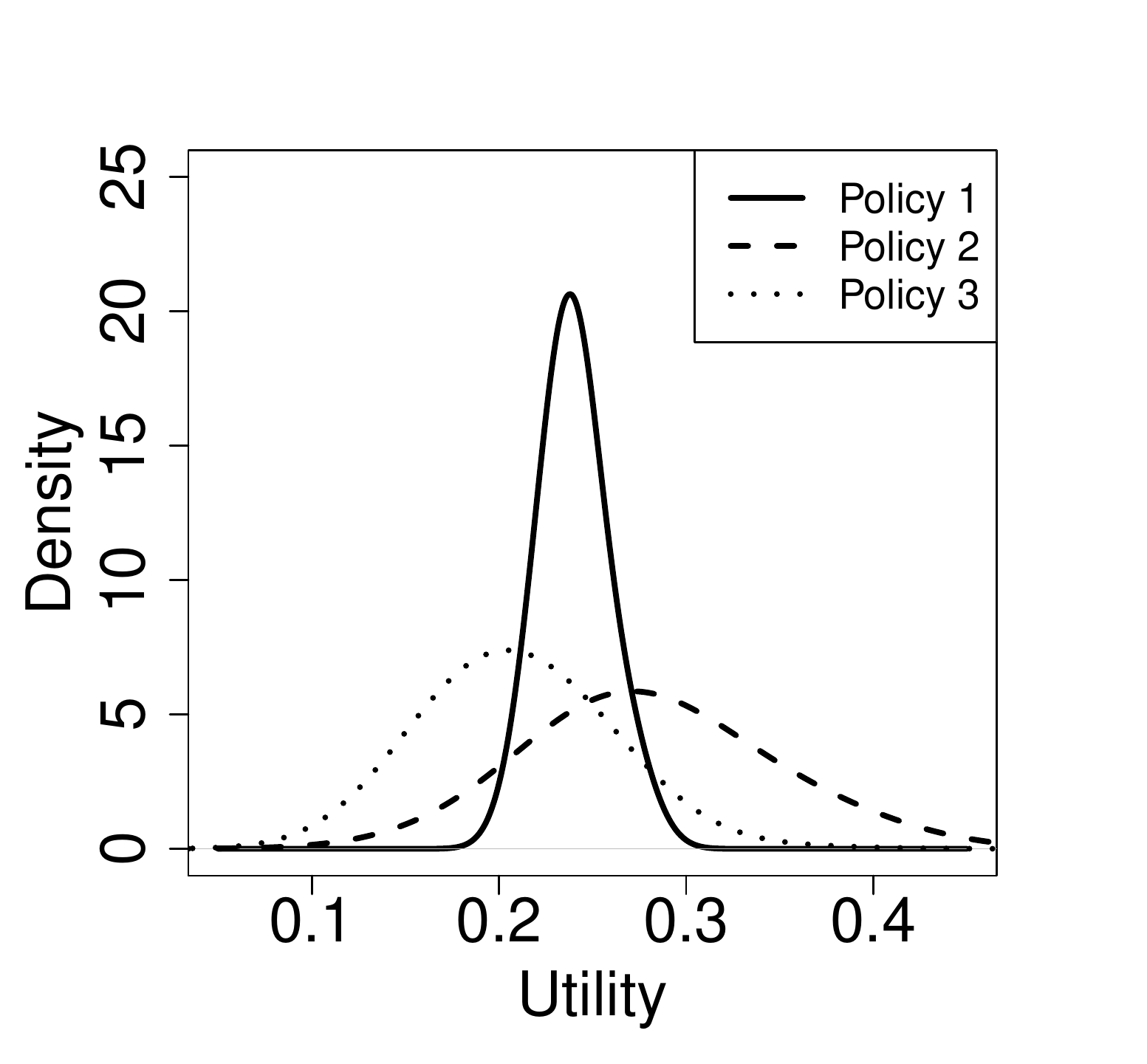} &
\includegraphics[width=5.8cm,height=5.2cm]{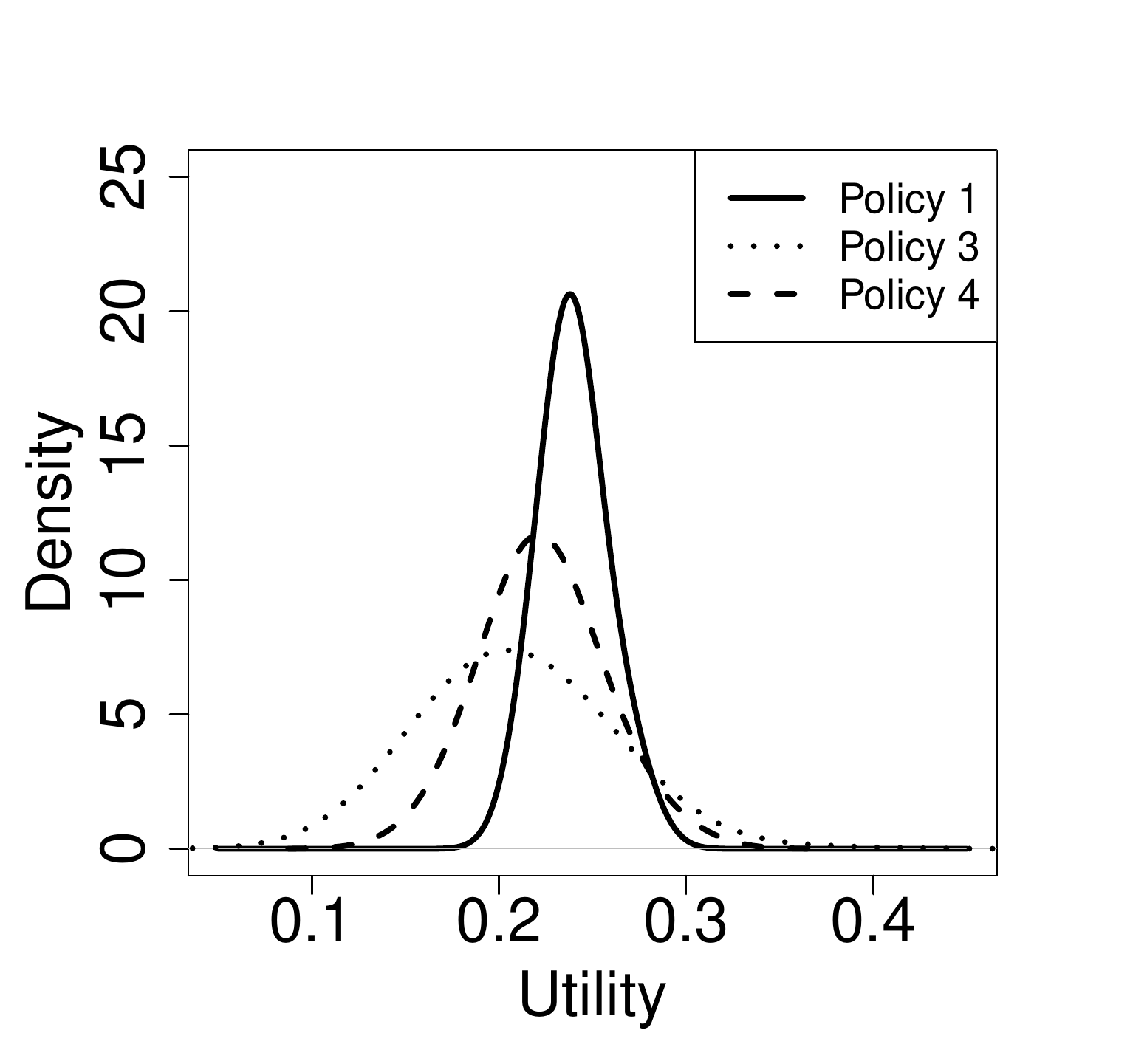} \\
\includegraphics[width=5.8cm,height=5.2cm]{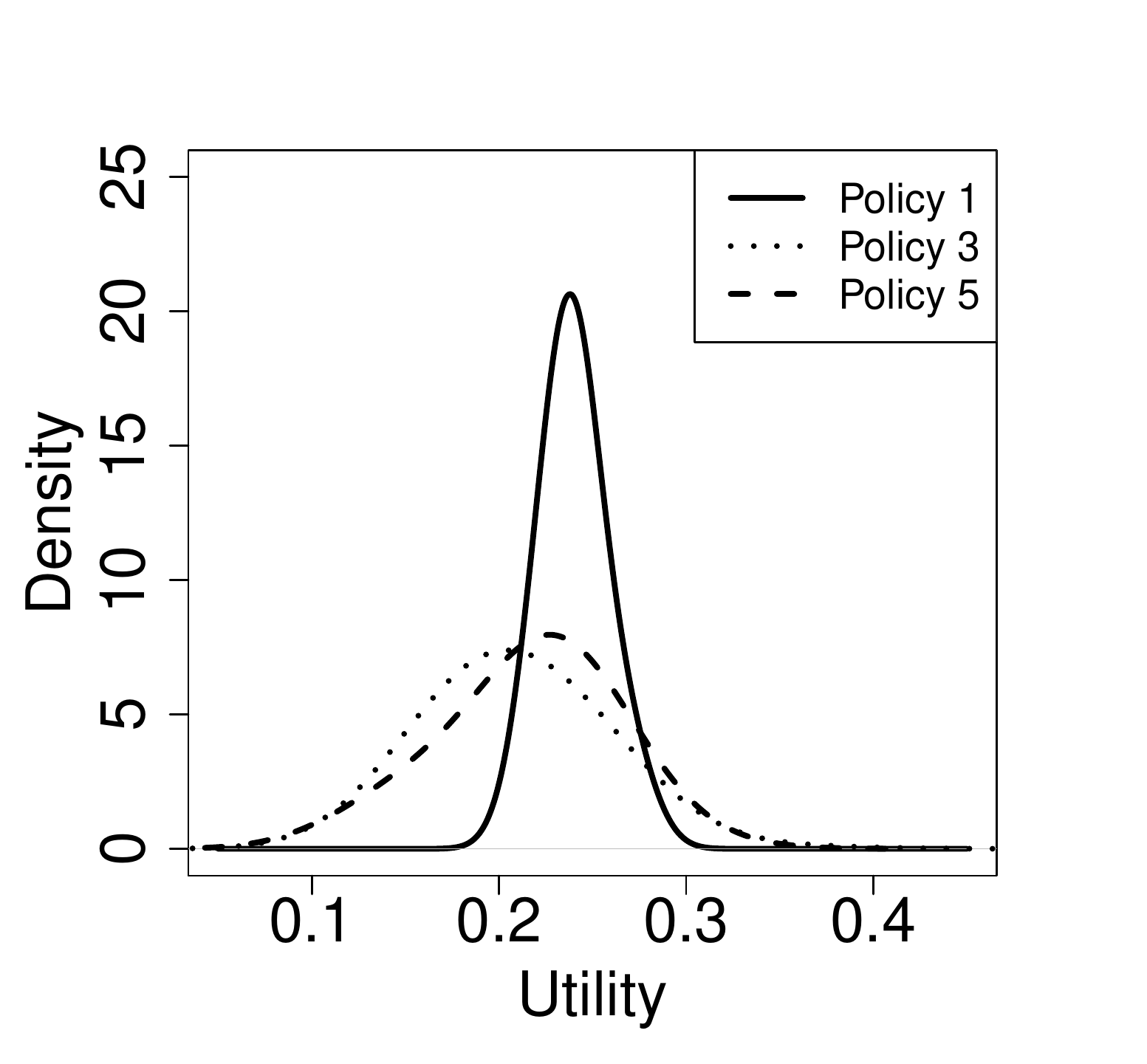} & \\
\end{tabular}
\caption{Utility function posterior distribution.}\label{Fig:u1}
\end{center}
\end{figure}

After fitting the dynamical model, different policies were compared using the IDSS approach described in section \ref{DDSS}. Policy 1 is  `do nothing', i.e.  all variables kept on the same observed values. Policy 2 accounts for an increase of $25\%$ in the food costs and policy, such as a no-deal Brexit \citep{Barons2020}. Policy 3 represents a decrease of $25\%$ in the food costs, such as through government subsidies. Figure \ref{Fig:u1} presents the posterior utility function for the 3 policies. Small values for the utility is associated with smaller rates of malnutrition and smaller percentage of disadvantaged pupils. The expected value of utility for policies 1, 2 and 3 are 0.2400, 0.2808 and 0.2091, respectively. Policy 4 considers the situation that food costs are reduced by $15\%$ plus household income is increased in $15\%$, through economic or welfare interventions. In this scenario the expected utility is 0.2232. Policy 5 is an agricultural policy leading to a reduced the output of food production (related to prices) by $25\%$ resulting in an expected utility of 0.2161. Note that the last scenario maintains the variables affecting food production as fixed in the observed values and modify the variables lower in the hierarchy such as food costs.

\newpage
\clearpage

\newpage
\clearpage

\section{Discussion and further developments}
We have shown a proof of concept IDSS for policymakers concerned with ameliorating household food security in the UK.  We have identified the main drivers of food security, drawing partly on research from the USA and Canada where food security has been measured for a number of years and therefore understanding of determinants of household food security are more advanced than in the UK.  We have identified plausible expert panels based on UK structures and have constructed models based on publicly available data.  We have demonstrated the output of the IDSS under a number of policies.  We have assumed equal weighting between health and educational attainment as a proxy for food insecurity. To move form a proof of concept to a working IDSS, we would need to elicit the user preferences for display of the results, as discussed in \citep{Barons2017}.

\newpage
\appendix

\section{Description of variables used in the network}

Panel G1 (household income) is represented by the variable HIncome. This variable depends on the household income after expenses.
\begin{itemize}
    \item[-] HIncome: Real net households adjusted disposable income per capita less the final consumption expenditure per head. 
\end{itemize} 
Panel G2 (food costs) is represented by the variable CFood. 
\begin{itemize}
    \item[-] CFood: CPI index of 9 food groups, 2015=100. Food costs was measured by a combination of CPI indices of items representing household dietary diversity \citep{Ken12}. The score is formed by 9 food groups: cereals, meat, fish, eggs, milk, oils and fat, fruits, vegetables and beverages. 
\end{itemize}
Panel G3 (income) accounts for access to credit (Lending), tax on the income (Tax), unemployment rate and social benefits. 
\begin{itemize}
    \item[-] Lending: Net lending (+)/net borrowing (-) by sector as a percentage of GDP - Household and non-profit institution serving households.
    \item[-] Tax: Tax on the income or profits of corporations.
    \item[-] Unemployment: Male unemployment rate, aged 16 and over, seasonally adjusted.
    \item[-] Benefits: Social assistance benefits in cash as a percentage of GDP.
\end{itemize} 
Panel G4 (costs of living) accounts for expenditure per head (Living) and housing costs (Chousing).
\begin{itemize}
\item[-] CLiving: Consumer price indices of main variables composing the expenditures of a household: Housing, including energy (CHousing), food (CFood), recreation (CRecreation),  and transport (CTransport).
\item[-] CHousing: CPI of housing, water and fuels. 
\end{itemize} 
Panel G5 (food supply) accounts for output of food production (FProduction) and imports from European Union and other countries. 
\begin{itemize}
    \item[-] FProduction: Output of food products.
    \item[-] FImports: Food imports from European Union countries plus imports from other countries.
\end{itemize} 
Panel G6 (Oil costs) is represented by CPI of fuels and energy (COil and CEnergy): 
\begin{itemize}
    \item[-] COil: Liquid fuels, vehicle fuels and lubricants (G) 2015=100.
    \item[-] CEnergy: CPI of energy, 2015=100.
\end{itemize}
Panel G7 (Demography) is represented by part-time work rates (PartTime). 
\begin{itemize}
    \item[-] Part-time: Part-time workers ( Ill or disabled).
\end{itemize}
Panel G8 (Weather) is represented by number of days in which the air temperature falls below 0 degrees Celsius. In these cases, sensitive crops can be injured, with significant effects on production. 
\begin{itemize}
    \item[-] Frost: Number of days of air frost.
\end{itemize}
Panel G9 (Economy) accounts for economic context represented by Gross D domestic Product (GDP): 
\begin{itemize}
    \item[-] GDP: Gross Domestic Product at market prices, seasonally adjusted.
\end{itemize}




\end{document}